\documentclass[useAMS,usenatbib]{mnras} %MNRAS
\usepackage{graphicx}
\usepackage{amssymb, amsmath}
\usepackage{times}
\usepackage{color}
\usepackage{todonotes}
\usepackage{lscape}
\usepackage[section]{placeins}
\usepackage{breqn}

\usepackage{lineno}

\usepackage{arydshln}
\usepackage{url}

\definecolor{carrotorange}{rgb}{0.93, 0.57, 0.13}

\newcommand\ihod{\texttt{iHOD}}

\newcommand{\gtsima}{$\; \buildrel > \over \sim \;$}
\newcommand{\ltsima}{$\; \buildrel < \over \sim \;$}
\newcommand{\simgt}{\lower.7ex\hbox{\gtsima}}
\newcommand{\simlt}{\lower.7ex\hbox{\ltsima}}

\def\mpcoh{\,h^{-1}{\rm Mpc}}

\begin{document}

%%%%% Front Matter %%%%%%%%%%%%%%%%%%%%%%%%%%%%%%%%%%%%%%%%%%%%%%%%%%%%%%%%%%%%

\title[Galaxy formation within the cosmic web]{Cosmic web dependence of galaxy clustering and quenching in SDSS}

\author[Alam et al.] {
    Shadab Alam$^{1}$ \thanks{ salam@roe.ac.uk}, Ying Zu$^{2}$\thanks{yingzu@sjtu.edu.cn}, John A. Peacock$^{1}$\thanks{jap@roe.ac.uk}, and Rachel Mandelbaum$^{3}$\thanks{rmandelb@andrew.cmu.edu}   \\
    $^{1}$ Institute for Astronomy, University of Edinburgh, Royal Observatory, Blackford Hill, Edinburgh, EH9 3HJ , UK \\
    $^{2}$ Department of Astronomy, Shanghai Jiao Tong University, 800 Dongchuan Road, Shanghai, 200240, China \\
    $^{3}$ Department of Physics and the McWilliams Center for Cosmology, Carnegie Mellon University, 5000 Forbes Ave., Pittsburgh, PA 15213, USA
}

\date{\today}
\pagerange{\pageref{firstpage}--\pageref{lastpage}}   \pubyear{2018}
\maketitle
\label{firstpage}

%\label{firstpage}
\begin{abstract}
    Galaxies exhibit different clustering and quenching properties in
    clusters, filaments, and the field, but it is still uncertain
    whether such differences are imprints of the tidal environment on
    galaxy formation, or if they reflect the variation of the
    underlying halo mass function across the cosmic web.  We measure
    the dependence of galaxy clustering and quenching on the cosmic
    web in the Sloan Digital Sky Survey, characterized by the
    combination of spherical overdensity $\delta_8$ and tidal
    anisotropy $\alpha_5$ centred on each galaxy. We find that galaxy
    clustering is a strong function of either $\delta_8$ or
    $\alpha_5$, and the large-scale galaxy bias shows complex and rich
    behaviour on the $\delta_8$ vs. $\alpha_5$ plane. Using the mean
    galaxy colour as a proxy for the average quenched level of
    galaxies, we find that galaxy quenching is primarily a function of
    $\delta_8$, with some subtle yet non-trivial dependence on
    $\alpha_5$ at fixed $\delta_8$. The quenched galaxies generally
    show stronger small-scale clustering than the star-forming ones at
    fixed $\delta_8$ or $\alpha_5$, while the characteristic scale at
    which the amplitude of clustering becomes comparable for both
    galaxy populations varies with $\delta_8$ and $\alpha_5$.  We
    compare these observed cosmic web dependences of galaxy clustering
    and quenching with a mock galaxy catalogue constructed from the
    \ihod\ model, which places quenched and star-forming galaxies
    inside dark matter haloes based on the stellar-to-halo mass
    relation and the halo quenching model --- the $\delta_8$ and
    $\alpha_5$ dependences of \ihod\ galaxies are thus solely derived
    from the cosmic web modulation of the halo mass function.  
    The main observed trends are accounted for extremely well by the \ihod\ model.  Thus any additional
    direct effect of the large-scale~(${>}5\mpcoh$) tidal field on
    galaxy formation must be extremely weak in comparison with the
    dominant indirect effects that arise from the environmental
    modulation of the halo mass function.
\end{abstract}

\begin{keywords}
    gravitation;
    galaxies: statistics, Quenching;
    large-scale structure of Universe;
\end{keywords}

%%%%% Main Text %%%%%%%%%%%%%%%%%%%%%%%%%%%%%%%%%%%%%%%%%%%%%%%%%%%%%%%%%%%%%%%
\section{Introduction}
\label{sec:intro}

The 3D distribution of galaxies in our Universe exhibits a
visually striking web-like pattern, consisting of dense knots of
galaxies as groups and clusters, connected by filaments and sheets,
with vast regions of cosmic voids in between~\citep{bond1996}. This
so-called `cosmic web' formed naturally out of the Gaussian initial
density field within $\Lambda$CDM~\citep{white1987}; this is a somewhat
surprising outcome considering that large-scale structure formed
hierarchically through the merging and accretion of dark matter
haloes~\citep{Peebles1980}, rather than the fragmentation of giant
`pancakes'~\citep{zeldovich1970, synyaev1972} --- an intuitively more
straightforward mechanism for generating sheet-like structures. One of
the key questions to ask is whether the properties of haloes and
galaxies depend on the local cosmic web environment.
Some such dependence is inevitable through the `peak-background split',
which states that the mass function of haloes is modified by the
large-scale density contrast \citep{Kaiser1984ApJ...284L...9K, ST1999};
thus it is inevitable that e.g. voids will be deficient in massive galaxies.
But beyond this leading-order effect, is there any evidence that
the cosmic web causes the galaxy population within it to change
with location? In other words, at a fixed density contrast, is there
any impact on the galaxy properties by the different tidal forces that are found in different
geometrical locations within the web? In principle, a hypothetical
effect of this sort might modulate the halo mass function beyond
the predictions of the peak-background split. But even if it does
not do so, it is possible that the mean galaxy content of a halo
of a given mass might depend on cosmic web location in addition to the halo mass. 
For example a massive halo close to a filament could slow down the growth of smaller halos in its vicinity~\citep{Hahn2009MNRAS.398.1742H, Borzyszkowski2017MNRAS.469..594B, Castorina2016arXiv161103619C} 
and a large scale tidal field alone can introduce a gradient in the accretion rate of halos \citep{Musso2018MNRAS.tmp..187M}.
Such cosmic-web effects are presumed
to be absent according to the widely-used
halo model framework~\citep{jing1998, ma2000, Seljak2000, Peacock2000,
  Benson2000, socco2001, Berlind2002, Cooray2002, yang2003,
  kravtsov2004, zheng2005, mandelbuam2006, leauthaud2011},
and the main aim of the present paper is to test this assumption.
We therefore measure the cosmic web dependence of the clustering and
quenching properties of galaxies in the Sloan Digital Sky
Survey~\citep[SDSS;][]{York2000}, and compare our results to the predictions
from a cosmic-web agnostic mock galaxy catalogue built from the
improved Halo Occupation Distribution model~\citep[\ihod;][]{zu2015,
  zu2016, zu2017}.

The clustering of dark matter haloes has an extremely complicated
dependence on the cosmic web environment. By measuring the halo-dark
matter cross-correlation and halo-halo auto-correlation functions
across four types of cosmic web environments (clusters, filaments,
sheets, and voids), \citet{yang2017} and \citet{Xia2017ApJ...848...22X} found that on scales larger than
$8\mpcoh$ the clustering bias of Milky Way-size haloes is significantly
enhanced in voids, but strongly suppressed in cluster
environments~(see their Figure 7). Using the SDSS main galaxy sample,
\citet{Abbas2007MNRAS.378..641A} also discovered a non-monotonic trend
of galaxy clustering with galaxy spherical overdensity, confirming the
theoretical expectation from the linear peaks bias
model~\citep{Kaiser1984ApJ...284L...9K, sheth1998}. However, they
found that the non-trivial scale and overdensity dependences of galaxy
clustering can be well reproduced by a simple three-parameter HOD
model that depends only on halo mass, suggesting that the observed
dependences largely stem from the density dependence of the halo
mass function~(HMF). We will extend the analysis of
\citet{Abbas2007MNRAS.378..641A} to measure the dependence of galaxy
clustering on the tidal anisotropy of the cosmic web environment, and
look for evidence of any deviation from the \ihod{} predictions. Such
a deviation would be direct evidence that the HOD of galaxies
depends not only on halo mass, but also on the properties of the
cosmic web --- a form of the so-called `galaxy assembly
bias'~\citep{croton2007, tinker2008, zu2008, wang2013, lin2016,
  zentner2016, tojeiro2017, guo2017, zehavi2017}.
Past searches for such effects have not reported any strong impact of the
anisotropic tidal field within the cosmic web on the mass
function of the dark matter haloes~\citep{Alonso2015MNRAS.447.2683A} or on
the luminosity function of galaxies~\citep{eardley2015}. However,
our work differs from these studies by using a different
measure of the tidal field, and we also focus on the
distinct question of galaxy quenching.

Halo mass is taken to be one of the main drivers of galaxy
quenching in theoretical models~\citep{dekel2006, cattaneo2006,
  correa2018}. Observationally, \citet{zu2016} found that the halo
quenching model provides significantly better fits to the clustering
and galaxy--galaxy lensing of SDSS galaxies than the other models that
tie quenching to stellar mass or halo age. The fiducial \ihod{} halo
quenching model also correctly predicts the average halo mass of the
red and blue centrals, showing excellent agreement with the direct
weak lensing measurements~\citep{mandelbaum2016}.  The \ihod{}
modelling of galaxy colours provides strong evidence that the physical
mechanism that quenches star formation in galaxies above stellar
masses of $10^{10}\,h^{-2}M_{\odot}$ is tied principally to the masses
of their dark matter haloes, rather than the properties of their
stellar components or halo age.

Besides halo mass, the cosmic web may also play a significant role in
the quenching of star formation in galaxies. In particular,
hydrodynamic simulations predict that the cold gas flow is often
directed along filaments~\citep{keres2005}, allowing galaxies to
efficiently accrete gas from large distances and sustain star
formation~\citep{kleiner2017}.  \citet{aragon-calvo2016} speculated
that the stripping of the filamentary web around galaxies is the
physical process responsible for quenching star formation, as cold gas
stripping, harassment, strangulation and starvation often happened
during the so-called `cosmic web detachment' events.  Therefore, it
is very important to understand the relative impact of the cosmic web on
galaxy quenching compared to that of halo mass~\citep{metuki2015}.

Recent studies of cosmic-web quenching mainly focused on the quenched
fraction of galaxies as a function of distance to the filaments. Using
the GAMA survey, \citet{kraljic2017} found that the red~(i.e.,
quenched) fraction increases when approaching knots and filaments, and
the star-forming population reddens at fixed mass when approaching
filaments~\citep[see also][]{guo2015,alpaslan2016, chen2017,
  kuutma2017, malavasi2017, poudel2017, odekon2017}.  In order to try
to disentangle the effect of overdensity from that of the anisotropic
tides, \citeauthor{kraljic2017} shuffled the colours of galaxies at
the same overdensity and stellar mass to preserve the colour-density-mass relation while
erasing any cosmic web effects. They found the cosmic web dependence
of galaxy colours substantially reduced in the reshuffled catalogue,
suggesting some unexplained tidal impact on galaxy quenching. However,
it is very challenging to disentangle further the effect of cosmic web
from that of halo mass using this shuffling technique, as the mass of
individual haloes is inaccessible from observations.

To distinguish between the cosmic web and halo mass effects on galaxy
formation, we build a mock galaxy catalogue by implementing the
\ihod\ prescription of \citet{zu2015} and the fiducial halo quenching
model of \citet{zu2016, zu2017}, without any explicit cosmic-web
effect in galaxy clustering or quenching. Any effect of the cosmic web on halo formation
and the halo mass function is implicitly taken into account by using haloes from N-body simulations. 
This \ihod\ halo quenching
model correctly reproduces the stellar mass and colour dependence of
galaxy abundance, spatial clustering, and galaxy-galaxy lensing, as
well as the conformity of galaxy colours~\citep{zu2017}, therefore
serving as an ideal cosmic-web agnostic benchmark model to compare
with observations --- any observed cosmic web dependences of galaxy
clustering and quenching that are unexplained by this \ihod\ halo
quenching model would provide important clues as to what aspect of
galaxy formation was shaped by the cosmic web. Similar \ihod\ halo
quenching mock catalogues were constructed by~\citet{zu2017b}
and~\citet{calderon2017} to study cluster assembly bias and galaxy
conformity, respectively.

The cosmic web is usually decomposed into four
structural elements --- knots, filaments, sheets, and voids, based on
the Hessian of the pseudo-gravitational potential computed from the
galaxy distribution ~\citep{Hahn2007MNRAS.375..489H}. Unfortunately,
the classification of any fixed galaxy position is not unique, but
depends on the choice of thresholds in the eigenvalues of the
Hessian~\citep{libeskind2018}. To circumvent this problem of
arbitrariness during the classification, we adopt the combination of
two orthogonal parameters: the spherical overdensity, and the `tidal
anisotropy' recently proposed by~\citet{paranjape2017} to characterize
the underlying cosmic web environment of each galaxy. Another
advantage of this method of characterizing the cosmic web is that the
two parameters are continuous variables, therefore allowing us to bin
galaxies arbitrarily on the 2D parameter plane.

The paper is organized as follows. We first introduce the SDSS data
and \ihod\ mock catalogues used in our analysis in section
\ref{sec:data}. In section \ref{sec:cweb} we describe the technique
used to compute galaxy overdensity and tidal anisotropy in the two
catalogues. We then discuss our results by comparing the observations
to \ihod\ predictions in section \ref{sec:result}.  We conclude by
summarizing our results and discussing their implications in section
\ref{sec:summary}. We assume a flat $\Lambda$CDM cosmology with
$\Omega_m{=}0.30$  unless specified otherwise. We also use $\lg$ to denote $\log_{10}$
everywhere in the text.

\section{Data and mock galaxy catalogues}
\label{sec:data}

In this section we briefly describe the data and mock galaxies used in this paper, including the selection of
the stellar mass-thresholded volume-limited galaxy catalogue from SDSS in Section~\ref{subsec:sdss}, and the
construction of the \ihod\ mock galaxy catalogue based on the halo quenching model in
Section~\ref{subsec:ihod}. For technical details in the making of the two catalogues, we refer readers to
\citet{zu2015, zu2016} for the SDSS data, and \citet{zu2017} for the \ihod\ mock, respectively.

\subsection{SDSS NYU-VAGC galaxy catalogue and sample selection}
\label{subsec:sdss}

We make use of the final data release of the SDSS I/II~\citep[DR7;][]{abazajian2009}, which contains the completed
data set of the SDSS-I and the SDSS-II. In particular, we obtain the Main Galaxy Sample~(MGS) data from the
\texttt{dr72} large--scale structure sample \texttt{bright0} of the `New York University Value Added
Catalogue'~(NYU--VAGC), constructed as described in~\citet{blanton2005}. We apply the `nearest-neighbour'
scheme to correct for the $7\%$ of galaxies that are without redshifts due to fibre collisions, and use data
exclusively within the contiguous area in the North Galactic Cap and regions with angular completeness greater
than $0.8$.

We employ the stellar mass estimates from the latest MPA/JHU value-added galaxy
catalogue\footnote{\url{http://home.strw.leidenuniv.nl/~jarle/SDSS/}}. The stellar masses were estimated based
on fits to the SDSS photometry following the methods of~\citet{kauffmann2003} and~\citet{salim2007}, and
assuming the Chabrier~\citep{chabrier2003} Initial Mass Function~(IMF) and the~\citet{bruzual2003} Stellar
Population Synthesis~(SPS) model.

For the purpose of our analysis, we select all the SDSS galaxies within the redshift range $z{=}[0.01,
0.074]$ and with stellar mass above $M_*^{\mathrm{min}}{=}10^{10}\,h^{-2}M_{\odot}$, yielding a sample of
$65222$ galaxies in total. Based on the stellar mass completeness limit estimated in \citet{zu2015}, we
believe this galaxy sample is roughly volume-complete down to $M_*^{\mathrm{min}}$, therefore can be used to
directly compare with any mock galaxy samples thresholded by $M_*^{\mathrm{min}}$.

\subsection{\ihod\ Mock galaxy catalogue with halo quenching}
\label{subsec:ihod}

We employ the `fiducial \ihod{} halo quenching mock' of \citet{zu2017} as our mock galaxy catalogue to
compare with the data. For the purpose of this paper, we use a cosmological dark matter-only \texttt{Bolshoi}~\citep{klypin11} simulation
because of its high dark matter mass resolution~($1.35\,{\times}\,10^8\,h^{-1}M_{\odot}$) and relatively large
volume~($250^3\,h^{-3}\mathrm{Mpc}^3$). We make use of the halo catalogues identified by the
\texttt{ROCKSTAR}~\citep{behroozi13} spherical overdensity halo finder at $z\,{=}\,0.1$. Finally, we populate
the halo catalogue with mock galaxies and assign them stellar mass and $g{-}r$ colours using the same
parameters as in \citet{zu2017}. We have implemented the best-fitting
velocity bias model of \citet{guo2015b}, which assumes that the relative velocities of central and satellite
galaxies follow Gaussian distributions within each halo.

In particular, the central galaxies and their host haloes follow a mean stellar-to-halo mass relation~(SHMR),
with a log-normal scatter that is also dependent on halo mass. Each halo also hosts a number of satellite
galaxies according to a conditional stellar mass function that only depends on the mass of that halo, with
their relative positions following an isotropic NFW distribution. Therefore, the stellar mass assignment in
the \ihod{} mock is tied only to the mass of host haloes, independent of the overdensity or tidal anisotropy of
the cosmic web environment.

To assign $g{-}r$ colours to the mock galaxies at fixed $M_*$ and $M_h$, we adopt the fiducial halo quenching
prescription described in \citet{zu2017}, so that the ranking order of galaxy colours at fixed $M_*$ depends
only on halo mass. The colour ranks are then converted to $g{-}r$ values according to the observed colour
distribution at that $M_*$. As a result, the quenching properties of the mock galaxies, as measured by the
simulated $g{-}r$ colours, also depend only on the halo mass.

\section{Galaxy environmental classification}
\label{sec:cweb}

In order to make a quantitative characterization of the cosmic web, we need to create a robust estimate of the tidal tensor field
using galaxies as tracers. We use the tidal tensor formalism developed
in~\citet{Hahn2007MNRAS.375..489H}~\citep[see also][]{Forero2009MNRAS.396.1815F, eardley2015}, and apply the
method to both the SDSS data and \ihod{} mock catalogues. In the rest of this Section we will describe the
technical details of the estimation method, but readers who are familiar with the method can skip
to~\S~\ref{sec:OTS} for the definition of $\delta_8$ and $\alpha_5$ based on the estimated tidal tensor field.

\subsection{Data regularization}

The first step of deriving the tidal tensor field is to solve the Poison equation, which can be done
efficiently in $k$-space using fast Fourier transforms. 
Naively, such procedure would
require placing the entire survey volume inside an encapsulating box. However, surveys like the SDSS usually
consist of a light-cone that extends to $200\mpcoh$ along the line-of-sight, with a large fraction of the
volume being empty. Therefore, to avoid having such a large box we regularize the survey volume and discard
galaxies outside our regularized volume for efficiency purposes. We first calculate the 3D Cartesian
coordinates for all galaxies assuming our fiducial cosmology, and examine the galaxy number
distribution along each of the three axes by using a histogram with 500 bins. For each axis, we then
find the threshold coordinate value by choosing the bins with number of galaxy $10\%$ of the bin
with highest number of galaxy. The cut-off threshold of $10\%$ is chosen empirically, so that we can
keep $98.5\%$ of the total galaxies while minimizing the volume of the encapsulating
box. This leaves us with a box of size 195$\times$340$\times$198 $\mpcoh$ with $65\%$ of the volume containing the galaxies, and the rest is empty. 

\subsection{Density estimation}
\label{sec:density}

After the regularization, we compute the local density field using a hybrid scheme based on the Voronoi
tessellation of the observed galaxy catalogue, for which an accompanying random galaxy catalogue has to be
generated to account for the survey mask and window function. Voronoi tessellation has been widely used in
modern cosmology for computing the galaxy density fields, providing a grid-free way of estimating
density~\citep{vandeWeygaert:2012:TAC:2261250.2261296} as opposed to count-in-cell methods. During the
tessellation the survey volume is partitioned such that the volume element closest to a given galaxy is
assigned to that galaxy, yielding a non-overlapping volume for each galaxy after combining all the volume
elements belong to that galaxy. We can then estimate the density at the location of each galaxy by computing
the inverse of the total volume associated with that galaxy.

However, in actual surveys the footprint usually has sharp boundaries, as well as holes inside the boundaries
that are masked out due to bright stars and/or bad seeing, so it could be difficult to account for the missing
area during tessellation. One approach to overcome this problem was proposed
by~\citet{VIDE2015A&C.....9....1S}, who associated the volume outside the immediate survey boundaries
with mock galaxies rather than any of the observed galaxies. This method is hard to implement, requiring the
generation of mock galaxies whose spatial distribution is statistically consistent with that of the observed
ones. Instead, we exploit the random galaxy catalogue that encodes the same angular and redshift selection
functions of the observed galaxies, and estimate density by directly counting the number of nearest random
galaxies around each observed galaxy.

For any galaxy near the survey boundaries or masks, we estimate the density at the location of that galaxy
using the inverse of the total number of closest randoms up to a constant multiplicative normalization factor.
This scheme inherits all the advantages of the traditional tessellation method, but is subject to sampling
noise of random points --- the noise in the density estimates depends on the total number of randoms used. In
general, such a method can be implemented very efficiently by using tree-based algorithms, which enable the use
of very large number of randoms, rendering the statistical noise negligible.

\subsection{Tidal tensor and eigenvalues}

After we estimate the local densities at the galaxy positions, we interpolate the density field on a regular
grid, which is then smoothed with a Gaussian kernel of width $5\mpcoh$. Ideally, the tidal anisotropy should
be evaluated at a few times the halo size~\citep{paranjape2017}, providing the redshift space smoothed density
$\delta_g^s$ on a grid.  We can then relate $\delta_g^s$ to the gravitational potential $\Phi$ through the Poisson
equation under the linear bias assumption
\begin{equation}
\nabla^2 \Phi = 4 \pi G \bar{\rho} \delta_g^s/b_g,
\end{equation}
where $G$ is the gravitational constant, $b_g$ is galaxy bias and $\bar{\rho}$ is the mean matter density. We note that above equation only applies in real space, but we ignore this and do not try to correct for redshift space. This is a reasonable assumption on large linear scales, which is why $5\mpcoh$ filtering is performed.
The above equation can be rewritten in terms of the dimensionless gravitational potential ($\tilde{\Phi}$):
\begin{equation}
\nabla^2 \tilde{\Phi} = \delta_g^s.
\label{eq:tphi}
\end{equation}
What we are really interested in is the dynamical evolution of two test particles~(in this case, two
galaxies) that are placed close to a given location.
In essence, the dynamics will be determined by the
relative gravitational acceleration between the two particles due to tidal force. The net force on a
particle is given by the negative gradient of the potential, while the relative acceleration is given by the gradient
of the force, which is the second derivative of the gravitational potential. In 3D space we can write the nine
components of this tidal force, i.e., the tidal tensor matrix $\tilde{T}^s$, as
\begin{equation}
 \tilde{T}^s= \begin{bmatrix}
 \partial^2 \tilde{\Phi}/ \partial x^2 & \partial^2 \tilde{\Phi}/ \partial x \partial y & \partial^2 \tilde{\Phi}/ \partial x \partial z \\[3pt]
 \partial^2 \tilde{\Phi}/ \partial y \partial x & \partial^2 \tilde{\Phi}/ \partial y^2  & \partial^2 \tilde{\Phi}/ \partial y \partial z   \\[3pt]
 \partial^2 \tilde{\Phi}/ \partial z \partial x & \partial^2 \tilde{\Phi}/ \partial z \partial y  & \partial^2 \tilde{\Phi}/
 \partial z^2  \end{bmatrix}.
\label{eq:Tmat}
\end{equation}
The above matrix describes the relative tidal force felt by an infinitesimally close pair of particles. To
classify galaxy locations into different geometric environments, we can simply count the number of positive
vs. negative eigenvalues of the local tidal tensor. The positive eigenvalues correspond to the directions
along which the particles will come closer, hence the collapsing direction, whereas the negative eigenvalues
represent the expanding directions. The tidal tensor matrix requires second order partial numerical
derivatives which are difficult to compute with noisy data, but we can use the standard Fourier
transform-based technique to solve for the components of the matrix in $k$-space. We first apply
Fourier transform to Equation~\eqref{eq:tphi} on both sides, yielding a simple solution for the tidal tensor
components in Fourier space:
\begin{equation}
\tilde{T}_{ij}^k =\frac{\partial^2 \tilde{\Phi}^k}{\partial_i \partial_j}=\frac{k_i k_j \delta_g^k}{k^2},
\end{equation}
where $\tilde{\Phi}^k$ and $\delta_g^k$ are the Fourier transforms of the gravitation potential and density
field, respectively.

In summary, to compute the matrix in Equation~\eqref{eq:Tmat}, we first estimate the density at each galaxy
location following the procedures described in~\ref{sec:density}. These densities are then interpolated onto a
regular grid of $5\mpcoh$ cell size, which can be smoothed with a Gaussian kernel to remove any discontinuities in the regions with missing data.
We use the fast Fourier transform (FFT) package in Python to transform the density field and estimate the
tidal tensor components $\tilde{T}_{ij}^k$. Finally, each of the tidal tensor components is estimated on a
grid, and then inverse Fourier transformed to obtain the tidal tensor in real space, as shown in
Equation~\eqref{eq:Tmat}.

\subsection{Overdensity and tidal shear}
\label{sec:OTS}

\begin{figure} \includegraphics[width=0.5\textwidth]{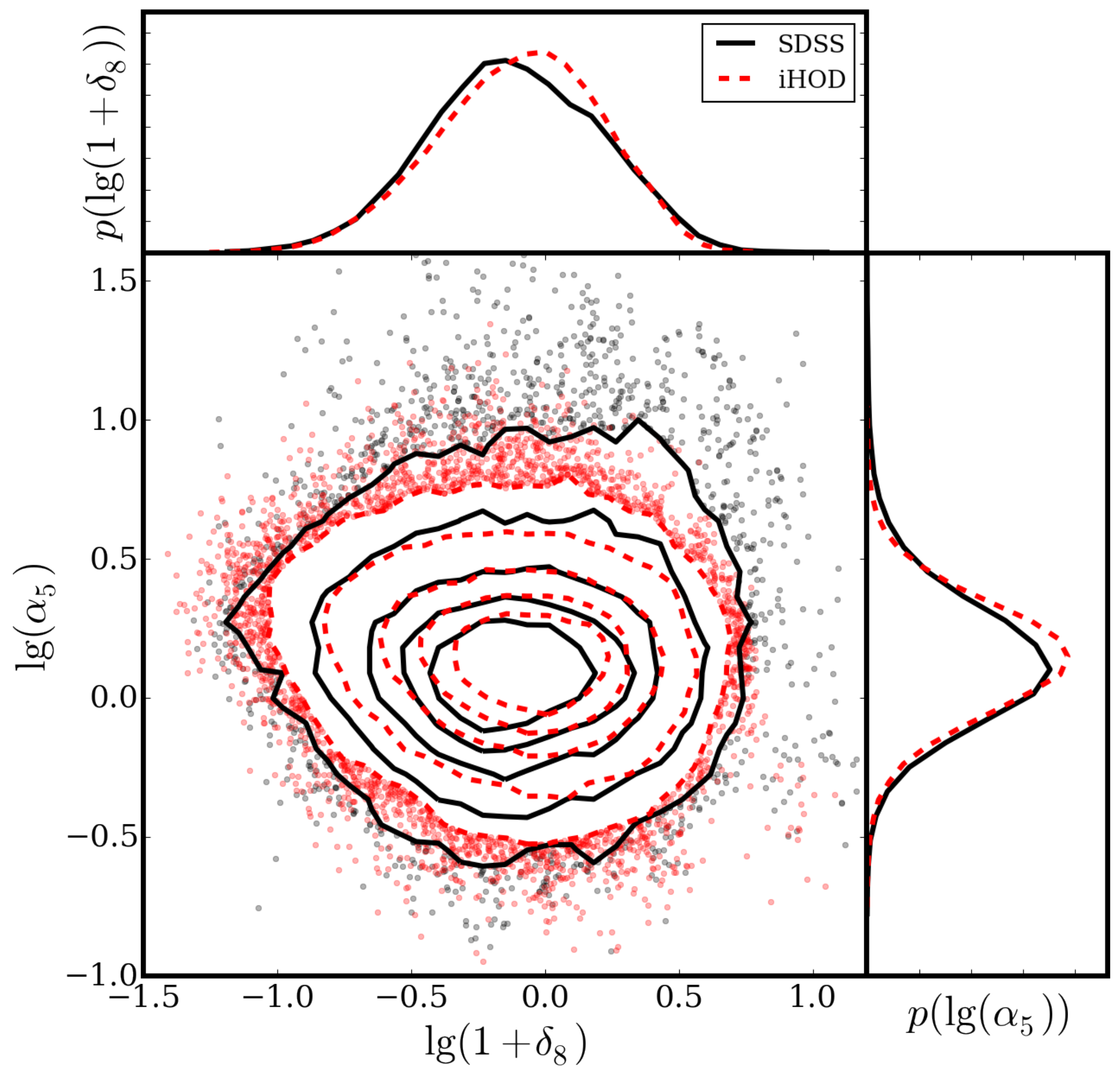}
    \caption{Comparison between the 2D distributions
	of SDSS~(black solid) and iHOD~(red dashed) galaxies on the $\delta_8$-$\alpha_5$ plane.
        The contour levels are 98\%, 90\%, 70\%, 50\%
        and 30\% inwards. The two sub-panels on the right and on the top show the marginalized 1D distributions of $\lg(\alpha_5)$ and
        $\lg(1+\delta_8)$, respectively. Note that the distribution of the iHOD galaxies matches that of the SDSS sample quite
        well, and by construction $\alpha_5$ is uncorrelated with $\delta_8$.}
    \label{fig:alpha-delta}
\end{figure}

As alluded to in the Introduction, we choose the combination of two continuous variables as a measure of the
cosmic web environment at each galaxy location. In particular, we define the first parameter $\delta_8$ as the
overdensity of galaxies enclosed within a sphere of radius $8\mpcoh$,
\begin{equation}
    \delta_8 = \frac{n_g^8 - \langle n_g\rangle}{\langle n_g\rangle},
    \label{eq:d8}
\end{equation}
where $n_g^8$ is the 3D number density of galaxies within the $8\mpcoh$-radius sphere and $\langle
n_g\rangle$ is the average galaxy number density of the sample.
We adopt the $8\mpcoh$ radius as it is large enough that the impact of small scale velocity dispersion (fingers of god) is negligible. But the effect of large scale `Kaiser anisotropy' will still exist. We choose to avoid any model-dependent correction for this and work with apparent quantities in redshift space that contain uncorrected Kaiser anisotropies.
To compute $\delta_8$, we first generate
$500$ random points inside a sphere of radius $8\mpcoh$ around each galaxy, and derive the densities at
those random locations by interpolating the grid values of the density field~(as calculated in
\S~\ref{sec:density}). We then obtain the local density $n_g^8$ by averaging over all the random points, and
compute $\delta_8$ from Equation~\eqref{eq:d8}.

\begin{figure*}
\includegraphics[width=1.0\textwidth]{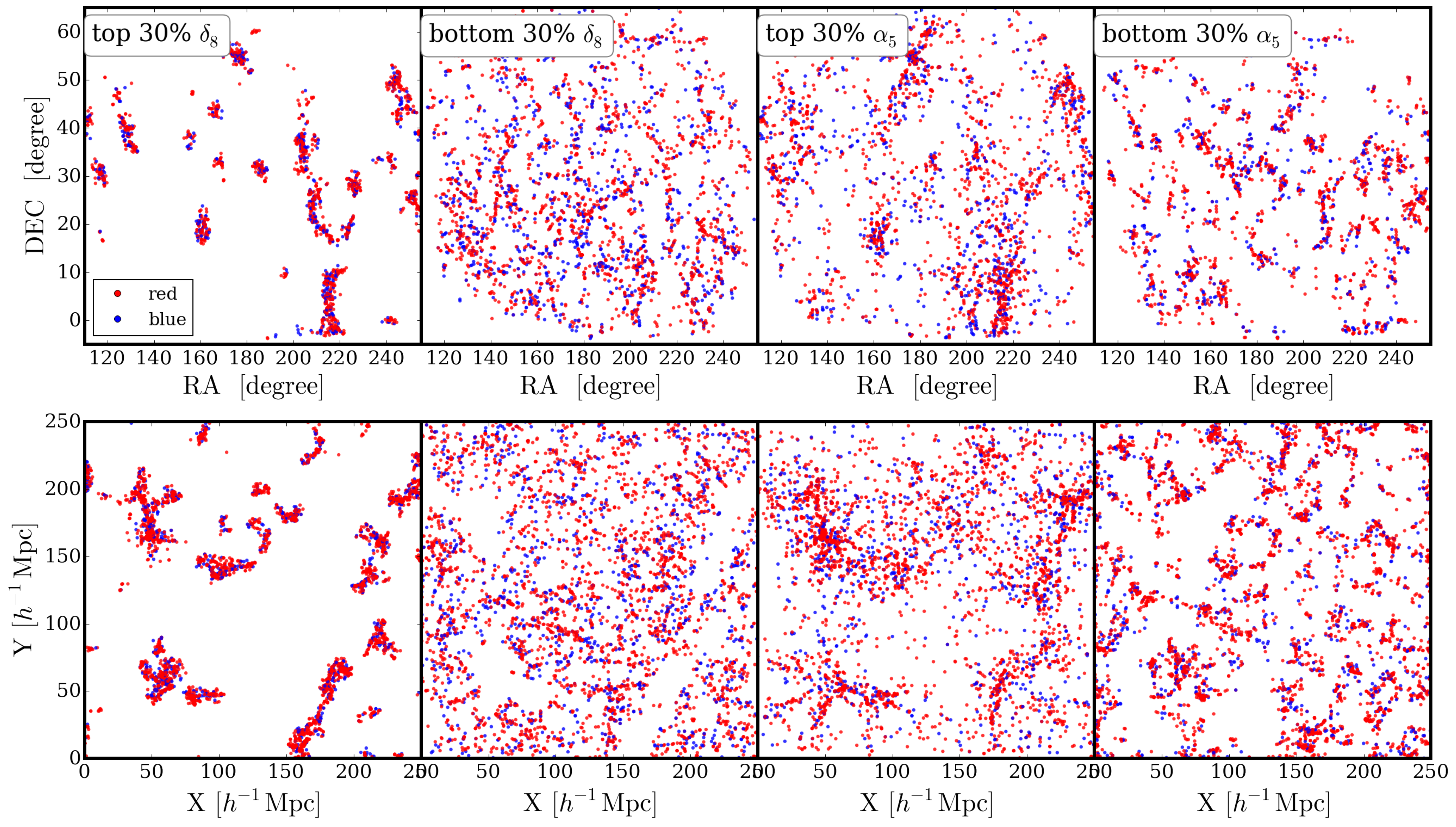}
\caption{The spatial distributions of SDSS~(top row) and iHOD~(bottom row) galaxies in the top/bottom $30\%$ of
    $\delta_8$~(left two columns) and $\alpha_5$~(right two columns), respectively.
    The red and blue points indicate the positions of galaxies with red and blue $g-r$ colours, respectively. The
    line-of-sight thickness of both samples is $20\mpcoh$. The RA/DEC ranges of the SDSS sample are chosen so that the
    SDSS galaxies subtend approximately $210\mpcoh$ at the median redshift~($z=0.057$) of the slice. 
    This figure demonstrates the visual similarity between the galaxy distributions in the  SDSS and
    the iHOD galaxies mock catalogues as a function of environment.}
\label{fig:xy-slice}
\end{figure*}

To complement $\delta_8$, we need a second parameter to describe the anisotropy level of the cosmic web, and
the tidal torque~($q^2_R$) is one such variable proposed in the
past~\citep{Heavens1988MNRAS.232..339H,Catelan1996MNRAS.282..436C}, defined as
\begin{equation}
    q^2_R=\frac{1}{2}[(\lambda_3-\lambda_2)^2+(\lambda_3-\lambda_1)^2+(\lambda_2-\lambda_1)^2],
    \label{eq:qr}
\end{equation}
where $\lambda_{1,2,3}$ are the eigenvalues of the tidal tensor defined in Equation~\eqref{eq:Tmat}, and we
adopt the convention that $\lambda_1<\lambda_2<\lambda_3$. The tidal torque is independent of $\delta_8$
for a Gaussian random field but not in the case of non-linear structure formation as we see in the real
universe~\citep[see section 3.2 of][]{paranjape2017}.  To remove the correlation between the tidal anisotropy
and overdensity, recently \citet{paranjape2017} proposed a new `tidal anisotropy' parameter $\alpha_R$,
which they defined as $\alpha_R{\equiv}\sqrt{q^2_R}(1+\delta_R)^{-n}$.  They adopted $n{=}1$ based on
empirical tests using simulations and set $R{=}4R_{200b}$, where $R_{200b}$ is the radius within which
enclosed matter density is $200$ times the background density. However, in real data we do not have $R_{200b}$
for each galaxy, so we use a constant smoothing scale $R{=}5\mpcoh$ in our analysis.  Additionally, the
correlation between $\delta_8$ and $\alpha_R$ depends on the smoothing scale and the bias of the galaxy
sample, so we adjust $n$ to be $0.55$ in the definition of $\alpha_R$, thereby eliminating any correlation
between $\delta_8$ and $\alpha_R$. The value of $n=0.55$ was determined by trying several values and looking for minimum correlation between $\alpha_5$ and $\delta_8$. In the end we have
\begin{equation}
    \alpha_5 = \sqrt{q^2_5}(1+\delta_5)^{-0.55},
    \label{eq:a5}
\end{equation}
where $\delta_5$ is the galaxy overdensity within a sphere of radius $5\mpcoh$ centred on each galaxy,
computed in the same way as $\delta_8$.  The value of $\alpha_5$ determines the spherical symmetry of the tidal and density field. Small $\alpha_5$ values correspond to isotropic regions containing voids and clusters. Intermediate values correspond to sheets and filaments and large values of $\alpha_5$ correspond to the filamentary (anisotropic) regions~\citep[see Figure 4 of][]{paranjape2017}.
We measure $\delta_8$ and $\alpha_5$ for each galaxy in the mock using the same methods as in the
data, so that any systematic uncertainties associated with the measurements~(e.g., the effect of the
smoothing scales and peculiar velocities) would not affect the results of our comparison between the
two catalogues. There can be some subtle residual difference between the data and mock due to
differences in cosmological parameters and strength of the Fingers of God (FoG) effect. The clusters
with a strong FoG effect can resemble a filamentary structure in redshift space and might lie in
high $\alpha_5$ region rather than the low $\alpha_5$ region corresponding to the isotropic
environment. If this effect is significant then we should see a discrepancy in both high and low $\alpha_5$
regions.

Figure~\ref{fig:alpha-delta} shows the joint distribution of $\delta_8$ and $\alpha_5$ for SDSS~(black solid)
and \ihod~(red dashed) galaxies. In the main panel, the contour levels are $98\%$, $90\%$, $70\%$, $50\%$, and
$30\%$ running inwards. The two subpanels on the right and top show the 1D marginalized distributions of
$\lg(\alpha_5)$ and $\lg(1+\delta_8)$, respectively. As expected,
$\alpha_5$ and $\delta_8$ show little correlation between each other, thereby providing almost orthogonal
information on the local cosmic-web environment. The overall distribution of the \ihod\ galaxies is in
good agreement with that of the SDSS galaxies on the 2D parameter plane, except in the high-$\delta_8$ and
high-$\alpha_5$ region where the mock galaxies are slightly under-populated. We selected these
outlier galaxies from data with $\lg(1+\delta_8)>0.3$ and $\lg(\alpha_5)>1$ where we found a very
small number of mock galaxies. We found that these galaxies are clustered together and lie between
redshifts of $0.06$ and $0.075$. We counted the number of galaxy groups and galaxy clusters from
NED\footnote{The catalogue of galaxy group and galaxy clusters between the redshift of 0.06 and
  0.075 were downloaded from \url{https://ned.ipac.caltech.edu/}}  around these outlier galaxies and found
that probability of finding more than 4 clusters within 1 degree is $0.70 \pm 0.09$. In contrast,
the probability of finding more than 4 clusters within a degree for a typical galaxy within the same
redshift is $0.25\pm 0.07$ and the probability of finding more than 4 clusters within a degree of
random points within the survey footprint is $0.07 \pm0.02$. The uncertainty on the probability
includes cosmic variance estimated based on N-body simulation by looking at the variance in the number of haloes above mass $10^{14} h^{-2}M_{\odot}$ in equivalent sub-volumes.
Therefore we suggest that the excess of
data galaxies with high-$\delta_8$ and high-$\alpha_5$ is associated with regions with large number of galaxy
clusters and galaxy groups probably forming a massive supercluster, a form of cosmic variance effect that is
difficult to capture in the \ihod{} mock due to its limited volume~\citep[but see][for an interesting example of constrained simulation to
resolve this mismatch.]{wang2016, yang2017b}.

\section{Results}
\label{sec:result}

\begin{figure*}
\includegraphics[width=1.0\textwidth]{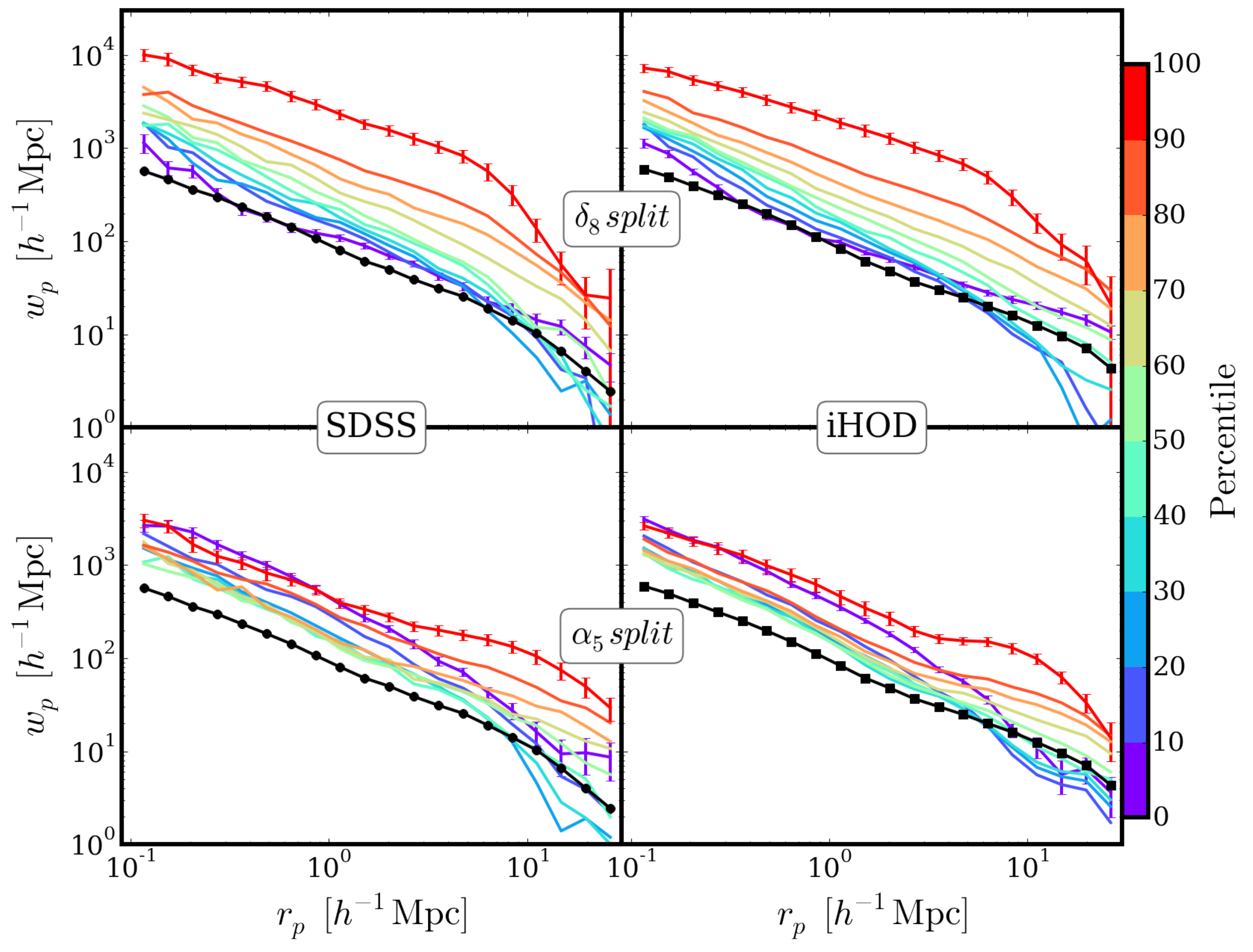}
\caption{The dependence of the projected auto-correlation function $w_p$ on $\delta_8$~(top row) and $\alpha_5$~(bottom row), for the
    SDSS~(left column) and \ihod{}~(right column) galaxies, respectively. In each panel, the black curve shows $w_p$ for the
    overall sample and the coloured lines show the auto-correlation $w_p$ in ten percentile bins of $\delta_8$ or
    $\alpha_5$ as indicated by the colourbar on the right. To avoid clutter, we only show error bars
    on the overall curve and the top/bottom $10\%$ subsamples in each split. Note that the error
    bars between different $r_p$ bins are correlated. This figure shows that $w_p$ is strong function of both $\delta_8$ and $\alpha_5$, and the behaviour is different at small and large scales. }
\label{fig:wp-10perc}
\end{figure*}

\begin{figure*}
\includegraphics[width=1.0\textwidth]{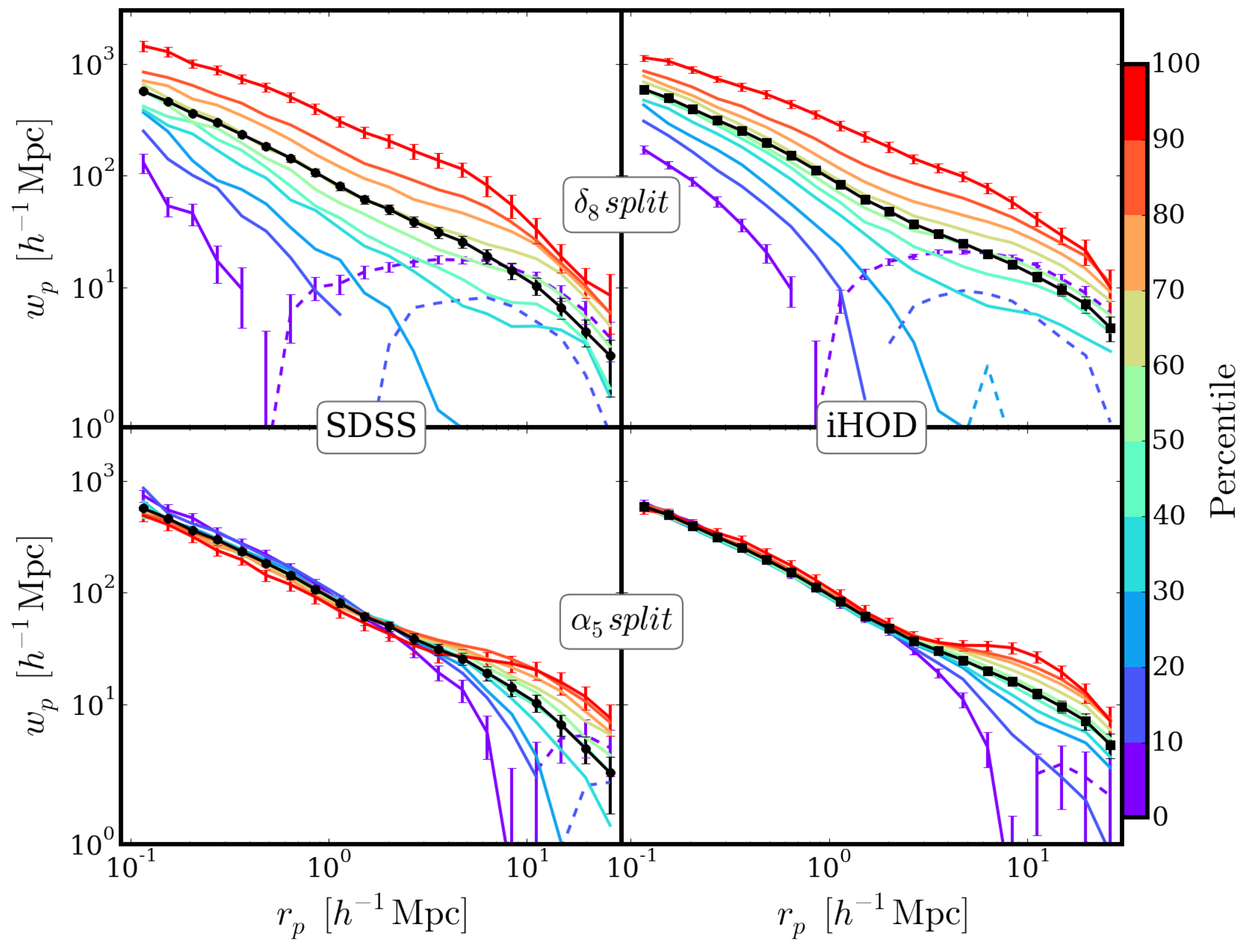}
\caption{The same as Figure \ref{fig:wp-10perc} but for
  cross-correlations of galaxies in selected environments with the
  full sample. The dashed lines indicate negative values where
  absolute values are plotted. This figure shows that $w_p$ is
  negative for low $\delta_8$ and $\alpha_5$ at large scales. This
  also shows that the small scale clustering is independent of
  $\alpha_5$ and hence is evidence of the fact that $\alpha_5$ is
  independent of $\delta_8$. }
\label{fig:wp-cross-10perc}
\end{figure*}

Due to the orthogonality between $\delta_8$ and $\alpha_5$, the combination of the two provides a compact
yet comprehensive measure of the cosmic web environment for each galaxy. In particular, for any two galaxies
at the same $\delta_8$, the one with the larger value of $\alpha_5$ is more likely to live in an anisotropic
environment like a filament or a sheet, while the one with the smaller value of $\alpha_5$ is more likely to
be found in a relatively isotropic environment, which is a knot if $\delta_8$ is high, or a void in the
low-$\delta_8$ case. Therefore, if galaxy clustering and quenching have extra dependence on the cosmic web
environment beyond halo mass, at fixed $\delta_8$ and $\alpha_8$ the SDSS galaxies should have a different
spatial distribution and red galaxy fraction than the \ihod{} mock galaxies.

Figure~\ref{fig:xy-slice} shows a visual comparison between the spatial distributions of red vs. blue galaxies
in four different types of environments within the SDSS~(top) and \ihod{}~(bottom) catalogues. In the top row, we
show the entire SDSS footprint within the northern galactic cap~(i.e., a comoving area approximately
$210\mpcoh{\times}210\mpcoh$) with a redshift range $\Delta z{=\,}0.007$ centred at
$0.057$~(${\simeq}\,20\mpcoh$).  For the \ihod{}
galaxies we select a slab of dimension $250\mpcoh{\times}250\mpcoh{\times}20\mpcoh$, slightly larger
than the SDSS footprint. The different columns show the top and bottom 30\% of galaxies in $\delta_8$ or
$\alpha_5$~(marked on the top left of each panel), and the red and blue points indicate galaxies with red and
blue $g-r$ colours, respectively. As we expected, the volumes occupied by the highest and lowest $\delta_8$
galaxies correspond to clusters/groups~(first column) and voids/fields~(second column), respectively.
Meanwhile, the top $30\%$ $\alpha_5$ galaxies trace the most prominent filamentary structures within the cosmic
web, while those with the bottom $30\%$ $\alpha_5$ have contributions from both the cluster regions and the
voids. Overall, the distributions of SDSS and \ihod{} galaxies exhibit very similar patterns when selected based
$\delta_8$ or $\alpha_5$, and we will compare the two in a quantitative fashion in the following sections.

\begin{figure*}
\includegraphics[width=1.0\textwidth]{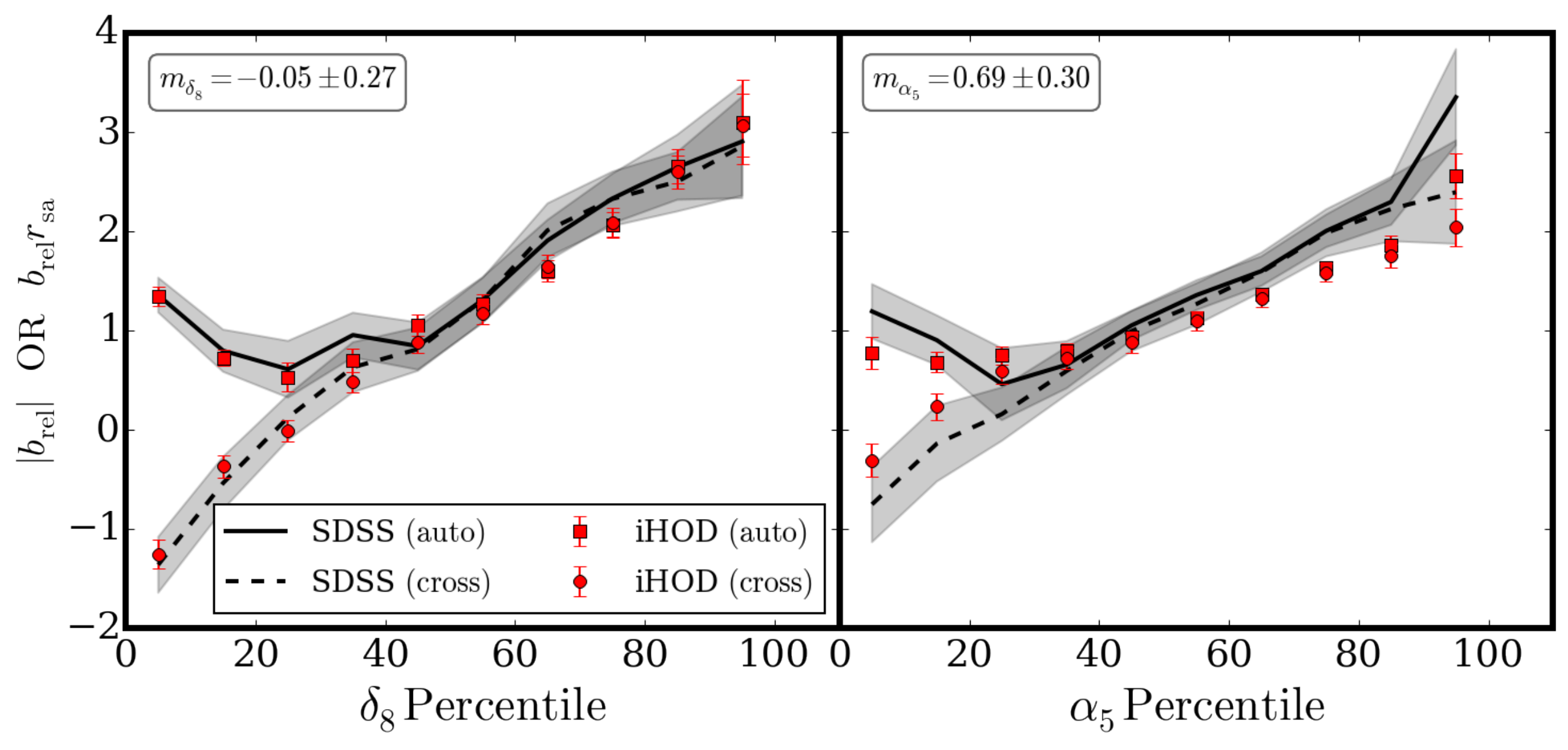}
\caption{The relative bias as a function of $\delta_8$~(left) or
    $\alpha_5$~(right), for the SDSS~(black curves with shaded uncertainty
    bands) and \ihod{}~(red points with error bars) galaxies, respectively. The
    relative bias  of the auto- (cross-)correlation is defined as the ratio between the  auto- (cross-)correlation $w_p$ of each subsample
    and that of the overall sample, averaged over scales between $10\mpcoh$ and $30\mpcoh$.
    The errors on the relative bias are estimated using jackknife sampling. Note that the error on
    the \ihod{} is smaller due to its larger volume compared to SDSS. The magnitude of the relative bias $|b_{\rm rel}|$ shows a
    characteristic dependence on both $\delta_8$ and $\alpha_5$, which is successfully
    reproduced by the \ihod{} model. The auto-correlation is only sensitive to the absolute value of
    the bias, but the cross-correlation shows that the bias is negative for low values of $\delta_8$
    or $\alpha_5$. We also show our linear model constraints on top left corner of each panel, where $m_{\delta_8}$ and $m_{\alpha_5}$ are slope parameters of our linear model given in Equation \ref{eq:lin-model} for the two environments respectively. }
\label{fig:bias-1d}
\end{figure*}

\subsection{Dependence of galaxy clustering on $\delta_8$ and $\alpha_5$}
\label{subsec:clustering}

We quantify the spatial clustering of galaxies using the projected correlation function $w_p$. To derive
$w_p$, the first step is to compute the 2D correlation function measured in bins of pair separations both
parallel~($r_\parallel$) and perpendicular~($r_p$) to the line-of-sight direction, using the Landy-Szalay
estimator~\citep{LandySzalay93},
\begin{equation}
\xi^{\rm LS}(r_p,r_\parallel)=\frac{DD(r_p,r_\parallel)-2DR(r_p,r_\parallel)+RR(r_p,r_\parallel)}{RR(r_p,r_\parallel)},
\label{eq:xi-LS}
\end{equation}
where $DD$, $DR$, and $RR$ represent the number counts of galaxy-galaxy pairs,
galaxy-random pairs, and random-random pairs, respectively. We use $20$
logarithmic bins in $r_p$~(between 0.1 and $30\mpcoh$) and $40$ linear bins in
$r_\parallel$~(between $-40$ and $40\mpcoh$). We then integrate $\xi^{\rm
    LS}(r_p,r_\parallel)$ along the line-of-sight to obtain $w_p$,
\begin{equation}
w_p = \int_{r_\parallel=-40}^{r_\parallel=40}  d r_\parallel \xi^{\rm LS}(r_p,r_\parallel).
\label{eq:wp}
\end{equation}
To minimize the impact due to the small discrepancies in the distributions of $\delta_8$ and/or $\alpha_5$
between the SDSS and \ihod{} catalogues, we always divide galaxies into subsamples based on quantiles instead
of fixed bin edges. To further remove the effect due to the residual correlation between the two parameters
when splitting galaxies by $\alpha_5$, we select fixed quantiles of $\alpha_5$ {\it at each ten percentile of
$\delta_8$}, and then combine the ten individual $\alpha_5$ quantiles to form the subsample at that
quantile. The errors on $w_p$ measurements are estimated using 200 jackknife realizations for data
and 100 jackknife realizations for the mock catalogue.

Figures~\ref{fig:wp-10perc} \& \ref{fig:wp-cross-10perc} compare the projected auto-correlations and cross-correlations of galaxies split by $\delta_8$~(top) and
$\alpha_5$~(bottom), between the SDSS~(left) and \ihod{}~(right) catalogues, respectively. In each panel, the
curves are colour-coded by the ten percentiles marked on the colourbar, ranging from the top $10\%$~(red) to
the bottom $10\%$~(purple), while the black circles indicate the $w_p$ of the overall sample. To avoid
clutter, we only show the 1-$\sigma$ measurement uncertainties for the top and bottom $10\%$ subsamples in
each panel. Clearly, the projected correlation function of the SDSS galaxies depends strongly on both
$\delta_8$ and $\alpha_5$, and the dependences on small scales are markedly different than on large scales.
Note that in general galaxies in regions of high $\delta_8$ belong to clusters/knots and those with low
$\delta_8$ live in the voids. Galaxies with low $\alpha_5$ usually reside in isotropic regions which could be
either clusters or voids, while those with high $\alpha_5$ reside in anisotropic tidal environment like sheets and
filaments.

When splitting by $\delta_8$~(top left), on scales below $r_p{\,\sim\,}2\mpcoh$ the observed amplitude of $w_p$
increases monotonically with $\delta_8$, while the $w_p$ shape stays largely unchanged. On scales above $2\mpcoh$, however, the amplitude of $w_p$ becomes a non-monotonic function of $\delta_8$, reaching its minimum
at the 30 percentile bin. Note that the clustering of the top 10\% $\delta_8$ bin decreases rapidly above
$r_p{\,\sim\,}8\mpcoh$, due to the selection effect imprinted by the characteristic scale that we used
to define $\delta_8$. One key feature of  cross-correlations (shown in Figure~\ref{fig:wp-cross-10perc}) compared to  auto-correlation (shown in Figure~\ref{fig:wp-10perc})  is that the clustering signal becomes negative at large scales for low $\delta_8$ sub-samples because galaxies are residing in under-dense regions, which is anti-correlated with the overall density field and hence appears as negatively biased regions. The overall dependence of $w_p$ on $\delta_8$ is consistent with the results of \citet{Abbas2007MNRAS.378..641A}. 
When splitting by $\alpha_5$~(lower left), the observed $w_p$ for auto-correlations are
non-monotonic functions of $\alpha_5$ on both small and large scales. In particular, the highest and
lowest 10\%-$\alpha_5$ subsamples exhibit similar levels of clustering on scales below $1\mpcoh$,
but their clustering strengths differ by almost a factor of five on scales larger than
$10\mpcoh$. Note that the measurement of auto-correlation for each subsample is higher than the
auto-correlation of the full sample below the smoothing scale used to estimate the environment: this
implies that there is a negative cross-correlation between the subsamples \citep{Abbas2007MNRAS.378..641A}. The observed $w_p$ for cross-correlations are independent of $\alpha_5$ at small scales below $5\mpcoh$, which is also our smoothing scale for $\alpha_5$. But at larger scales they show a strong non-monotonic dependence on $\alpha_5$, and the lowest sub-samples of $\alpha_5$ have negative cross-correlations at large scales (shown with dashed lines in Figure \ref{fig:wp-cross-10perc}).

Such rich and complex phenomenologies in the dependence of $w_p$ on $\delta_8$ and $\alpha_5$, however, are
closely reproduced by the \ihod{} mock galaxies in the two right panels. Since the apparent cosmic web
dependence of galaxies in the \ihod{} mock is entirely caused by the variation of halo mass function with
large-scale environment, the good agreement between the left and right columns in Figures~\ref{fig:wp-10perc} \& \ref{fig:wp-cross-10perc} suggests that the cosmic web dependence of clustering in the SDSS can be largely explained by an \ihod{} mock that does not include any galaxy assembly bias, and that any direct impact of the cosmic web on galaxy clustering must be weak in comparison.

To highlight the cosmic web dependence of galaxy clustering in the linear regime, we also measure the
large-scale relative bias $b_{\mathrm{rel}}$ as functions of $\delta_8$~(left) and $\alpha_5$~(right), as
shown in Figure~\ref{fig:bias-1d}.  Throughout the paper, we define the {\it relative bias} of a subsample as
the ratio between the bias of either auto or cross correlations of that subsample and that of the overall sample. 
The auto- and cross-correlations of subsamples are related to the auto-correlation of all galaxies as follows:

\begin{align}
w_p^{\rm ss} &= b^2_{\rm rel} w_p^{\rm aa}  \\
w_p^{\rm sa} &= b_{\rm rel} r_{\rm sa} w_p^{\rm aa},
\end{align}
where $w_p^{\rm aa} ,w_p^{\rm sa}$ and $w_p^{\rm ss}$ represent the projected correlation functions for all galaxies, cross-correlations of subsamples with all galaxies and auto-correlations of subsamples, respectively. Note that based on this definition, the auto-correlation measurements only provide the magnitude of relative bias and the cross-correlation measurements reveal the product of the relative bias with the cross-correlation coefficient ($r_{\rm sa}$). We use jackknife resampling in order to account for the covariance of $w_p$ between different
distance bins including the covariance between the subsample and overall sample. We first estimate the
error on the $w_p$ measurements for the subsample and the overall sample using the variance of jackknife realizations.  For each jackknife realization, we estimate the value of relative bias ($b_{\rm rel}$) corresponding to minimum $\chi^2$ between projected clustering of a subsample ($w_p^{\rm sub}$) and overall sample ($w_p^{\rm aa}$).
The $\chi^2$ is given by the following equation:
\begin{equation}
\chi^2= \sum_{r_p \geq 10}^{r_p<30}{ \left( \frac{w^{\rm sub}_p(r_p)-b_1 b_2 w^{\rm aa}_p(r_p)}{\sigma_{w_p}(r_p)}\right)^2},
\end{equation}
where $w^{\rm sub}_p$ represents $w^{\rm ss}_p$ for auto-correlation and $w^{\rm sa}_p$ for cross-correlations. The diagonal error on $w_p$ is estimated by adding the error of subsample and overall sample in quadrature 
\smash{($ \mathrm{\it{i.e.}} \, \sigma_{w_p}^2=\sigma_{w_p^{\rm sub}}^2+\sigma_{w_p^{\rm aa}}^2$)}.
For auto-correlations $b_1=b_2=b_{\rm rel} $ and for cross-correlations $b_1=b_{\rm rel}$  and $b_2=r_{\rm sa}$. We then compute the mean relative bias and its uncertainty using the bias measurements from all of the jackknife realizations.
In each panel, the solid and dashed black line with shaded band indicates the relative bias function and its 1-$\sigma$
uncertainties measured from the auto and cross correlations of SDSS galaxies, while the red squares and circles  with error bars are from the \ihod{} mock galaxies auto and cross correlations respectively. 
In the left panel, the observed relative bias function from the cross-correlation monotonically increases, with a zero-crossing around the $20{-}30$ percentile bin and a positive slope. This also shows a characteristic asymmetry where the highest $\delta_8$ bin has a stronger magnitude of bias compared to the lowest $\delta_8$ bin. 
This behaviour of the relative bias can be understood in the context of the
peak background split formalism~\citep{Kaiser1984ApJ...284L...9K, ST1999} as follows: in the initial Gaussian
random field, the highest and the lowest $\delta_8$ regions are equally rare, and therefore should both have
similarly high biases. The asymmetry emerged as structures formed non-linearly at later epochs, when the
$\delta_8$ distribution developed a non-Gaussian tail into the high overdensity regime, hence the enhanced
magnitude of the bias relative to the under-dense regions. The negative sign of the bias for low $\delta_8$ regions is expected, as the galaxies lying in regions with large-scale density below the mean density will be anti-correlated with the matter field and show negative bias. The auto-correlation function shows essentially the same behaviour except that the auto-correlation is only sensitive to the magnitude of the bias and hence it is positive even in low $\delta_8$ regions. 
The amplitude and shape of the observed $b_{\mathrm{rel}}(\delta_8)$ function are accurately reproduced by the \ihod{}\ mock galaxies,
indicating that the $\delta_8$-dependence of galaxy clustering can be entirely attributed to the overdensity-dependence of the underlying halo mass function. Our findings here for the auto-correlations are consistent with the results of~\citet{Abbas2007MNRAS.378..641A}. We also note that the magnitude of the bias from auto-correlations and cross-correlations are very close to each other for all values of $\delta_8$ and hence the cross-correlation coefficient $r_{\rm sa}(\delta_8)$ must be very close to 1 for this sample.

Intriguingly, the observed $\alpha_5$-dependence of $b_{\mathrm{rel}}$ from cross-correlations also exhibits a negative bias in low $\alpha_5$ bins with an asymmetric magnitude in low- and high-$\alpha_5$ bins in
the right panel, with the zero crossing occurring at the $10{-}20$ percentile of $\alpha_5$. 
Despite the similarities, this $\alpha_5$-dependence of the relative bias is independent of $\delta_8$, i.e., the
large-scale galaxy clustering is an intrinsically strong function of $\alpha_5$ regardless of $\delta_8$, as
the conditional probability distributions of $\delta_8$ are uniform across different $\alpha_5$
percentiles~(by design).  The $\alpha_5$ dependence predicted by the \ihod{}\ model is qualitatively consistent
with the SDSS measurements, except at high $\alpha_5$ where the predicted relative bias is slightly lower
than the observations. In addition, the 1D magnitude of bias in the mock does not rise as significantly as in the data
towards the lowest $10\%$ bin, but is nonetheless consistent with the observations within the uncertainties. Similar to $\delta_8$, the magnitude of the relative bias from auto-correlations and cross-correlations is very similar as a function of $\alpha_5$ except in the lowest and highest bins, and hence the cross-correlation coefficient $r_{\rm sa}(\alpha_5)$ must be very close to 1 for this sample except for extreme values of $\alpha_5$ for both SDSS and \ihod.

It is interesting to consider the formal degree of agreement between the above measurements and
the \ihod{}\ predictions. This can be addressed directly by using our jackknife realizations
to estimate the covariance matrix for the bias data. Considering both the 
auto- and cross-correlation estimates together, we can evaluate $\chi^2$ on 20 degrees
of freedom. The  $\delta_8$ measurements yield $\chi^2=8.4$, which is statistically
acceptable; the $\alpha_5$ results agree less well visually but statistically give $\chi^2=25.2$ -- corresponding
to a $p$-value of 20\%.  If one regards the current outcome as a null result, it is
interesting to ask what level of effect could be clearly detected
by an analysis of this sort.  
As an illustrative answer to this question, we have considered an empirical linear model
for the relation between relative bias and environment: 

\begin{equation}
b_{\rm rel}^{\rm SDSS} ({\rm env}) r_{\rm env} =  b_{\rm rel}^{\rm \ihod{}} ({\rm env})r_{\rm env} + m_{ {\rm env}} R_{{\rm env}} + c_{{\rm env}}\, ,
\label{eq:lin-model}
\end{equation}
where ${\rm env}$ is either $\delta_8$ or $\alpha_5$ with $m_{\rm env}$ and $c_{\rm env}$ are the two parameters of our
model accounting for any systematic difference between SDSS and
\ihod{}. Here, $R_{\rm env}$ represents the rank of ${\rm env}$ bins
mapped to the interval [$-$0.5,\,0.5]. In this model, $b$ means the bias
with definite sign inferred from cross-correlation, so it is not a
linear model for $|b|$. This means for auto-correlations $r_{\rm env}$ is set to $+1$ for bins above $20{-}30$ percentile and $-1$ for lower percentile bins whereas for cross-correlations we directly measure the product $b_{\rm rel} r_{\rm env}$.
We ran an MCMC analysis to
estimate $m_{\rm env}$ after marginalizing over $c_{\rm env}$ and
accounting for covariances in the measurement of relative bias using
the jackknife. The results show that $m_{\alpha_5}$ can be
constrained with an rms error of 0.30, so that effects with slopes
above about 1 could have been clearly detected if they had been present. 
We also found that for \ihod{} the $m_{\alpha_5}=0.69 \pm 0.30$ which is $2.3\sigma$ deviation. 
It is a matter of debate whether this degree of mismatch should
be taken as evidence for a physical effect of the cosmic web environment, or whether it
might indicate a small imprecision in the \ihod{}\ mocks. For example, our measurements
are made in redshift space and are thus sensitive in particular to the exact amplitude
of Finger-of-God virialized velocities. We have used N-body simulations to study the impact of peculiar velocity on such measurements and found that a similar deviation in $m_{\alpha_5}$  can be produced when the peculiar velocities of galaxies are scaled by a factor of 3. Exploring 
marginalization over such effects is beyond the scope of this work, but because of the outcome of this simple peculiar velocity study,  we do not currently
claim any strong evidence for detection of direct tidal effects of the cosmic web.
 We have also measured $m_{\delta_8}=-0.05 \pm 0.27$ , which is statistically consistent with zero. 
So, if one considers both $m_{\alpha_5}$ and $m_{\delta_8}$ together then we find no strong evidence for a deviation between SDSS and \ihod{} using this simple linear model.

\begin{figure*}
\includegraphics[width=0.98\textwidth]{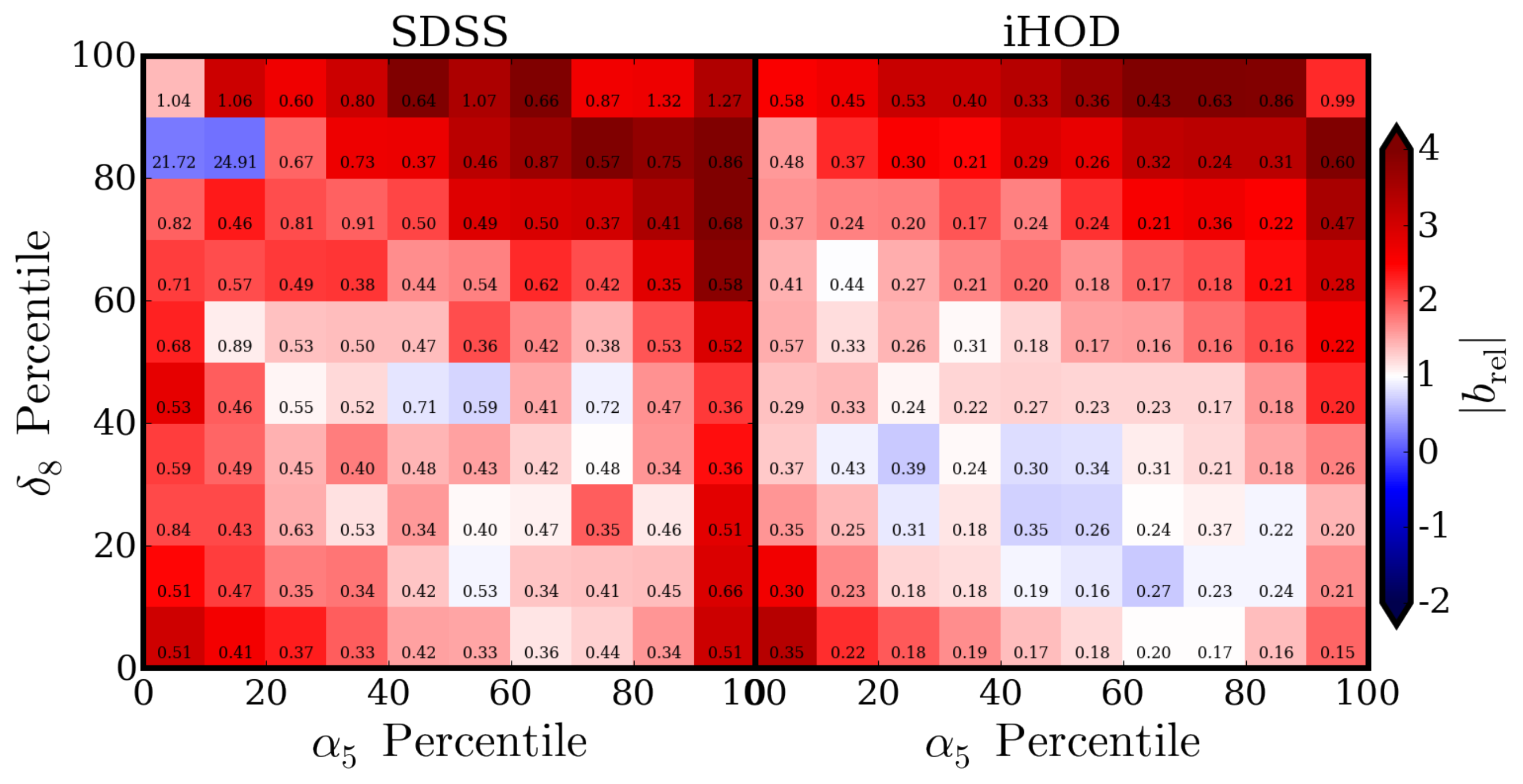}
\caption{Comparison of the relative biases between the SDSS~(left) and
\ihod{}~(right) galaxies on the $\delta_8$-$\alpha_5$ plane.
    Each cell is colour-coded by the bias relative to the overall sample of
    galaxies in a 2D bin of ten percentile in $\delta_8$ and $\alpha_5$ jointly. The numbers
    displayed in each cell show the statistical errors on the relative bias of the corresponding cell.
}
\label{fig:bias-2d}
\end{figure*}

\begin{figure*}
\includegraphics[width=0.98\textwidth]{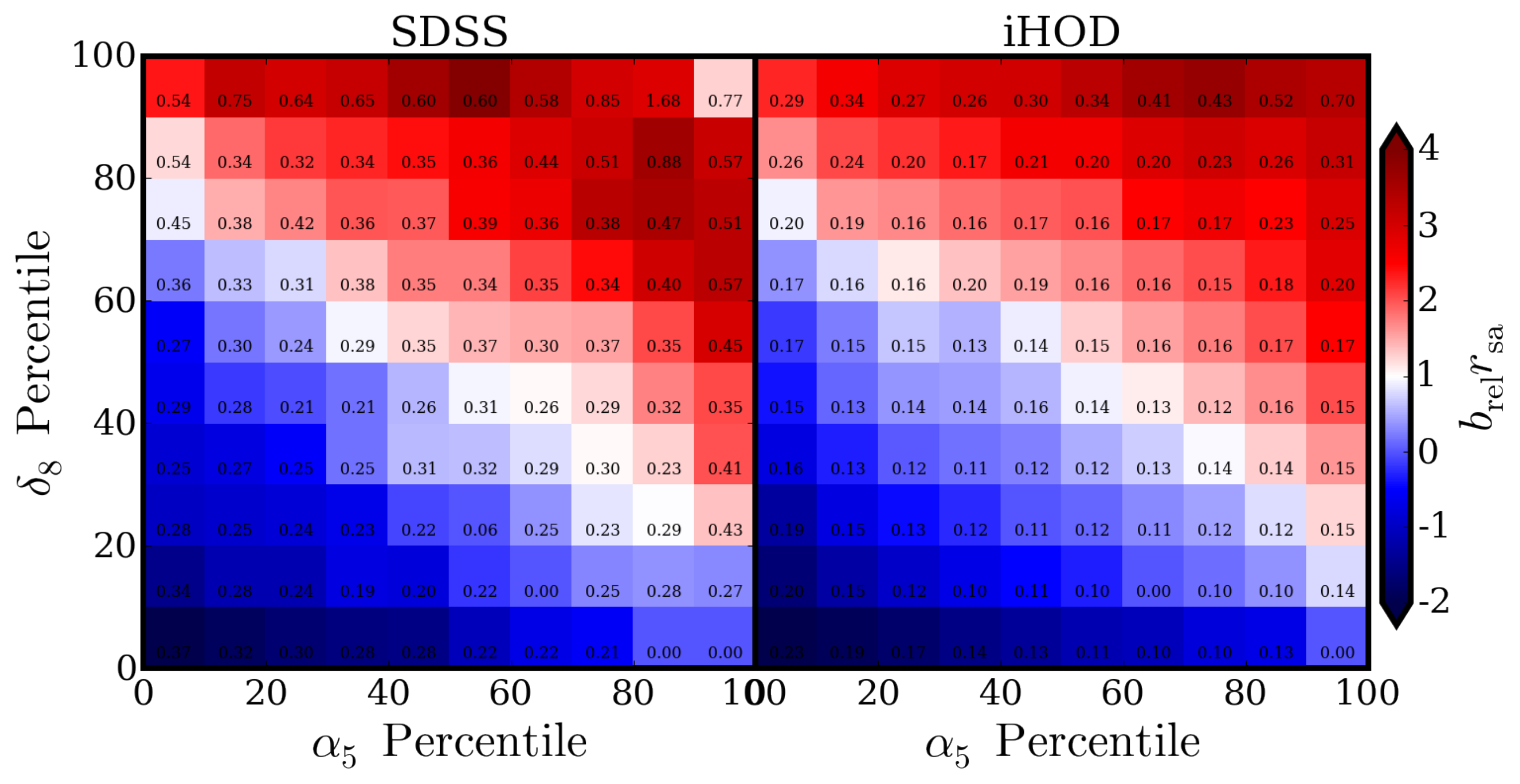}
\caption{ Similar to Figure \ref{fig:bias-2d} but using cross-correlation to measure relative bias.
}
\label{fig:bias-2d-cross}
\end{figure*}

To better understand the origin of the strong dependence of $|b_{\mathrm{rel}}|$ with $\alpha_5$ in both
the data and mock catalogues, we can examine the relative bias as a 2D function of both variables
simultaneously, by calculating the projected auto-correlation functions of subsamples defined on the
$\delta_8$ vs. $\alpha_5$ plane.  Figure~\ref{fig:bias-2d} compares this 2D bias function measured from
SDSS~(left) to that from the \ihod{}\ mock~(right). In each panel, the colour-coding indicates the value of the
relative bias in each cell of ($\alpha_5$, $\delta_8$), with the colour bar shown on the right.  The numbers displayed in each cell shows the errors on the relative bias of the cell. Similarly Figure~\ref{fig:bias-2d-cross} shows the 2D bias function but measured using cross-correlations. It is clear
that the observed galaxy bias depends on both $\alpha_5$ and $\delta_8$ simultaneously, and the 2D bias map
shows several interesting features on the $\alpha_5$ vs. $\delta_8$ plane:
\begin{itemize}
    \item The low-bias galaxies live predominantly in under-dense regions~(bottom $50\%$ $\delta_8$), but at
	each fixed $\delta_8$ percentile, the lowest magnitude of bias occurs at different percentiles of $\alpha_5$ ---
	galaxies with high $\alpha_5$ and low $\delta_8$ have a similar bias to those with low $\alpha_5$ and
	intermediate $\delta_8$.
    \item Galaxies that live inside the densest portion of the filaments~(high-$\delta_8$ and
	high-$\alpha_5$; top right corner) have the strongest large-scale clustering bias.
    \item Galaxies that live in isolated voids~(low-$\delta_8$ and
	low-$\alpha_5$; bottom left corner) have a relatively strong bias compared to those in the field but with negative sign.
    \item Galaxies that live in the most isolated and densest knots~(top $10\%$
	$\delta_8$ and bottom $10\%$ $\alpha_5$; top left corner) have a
	relatively low bias, reflecting the rapid decline of $w_p$ on scales above $8\mpcoh$~(see
	the red curve in the top left panel of Figure~\ref{fig:wp-10perc}).
\end{itemize}
Moreover, we can better understand the dependences of the 1D relative bias~(seen in
Figure~\ref{fig:bias-1d}) by averaging the 2D bias map along each of the two axes.  Specifically, the
strongest 1D bias at the high $\alpha_5$ or high $\delta_8$ end is contributed solely by galaxies inside the
dense filaments, while the enhanced 1D bias at the bottom $10\%$ of $\alpha_5$
or $\delta_8$ is caused by galaxies inside the isolated voids. The minimum 1D bias occurs at the $20{-}30$
percentile of $\alpha_5$ or $\delta_8$, primarily due to the lack of any such galaxies living in dense
filaments or isolated voids.

\begin{figure*}
\includegraphics[width=0.8\textwidth]{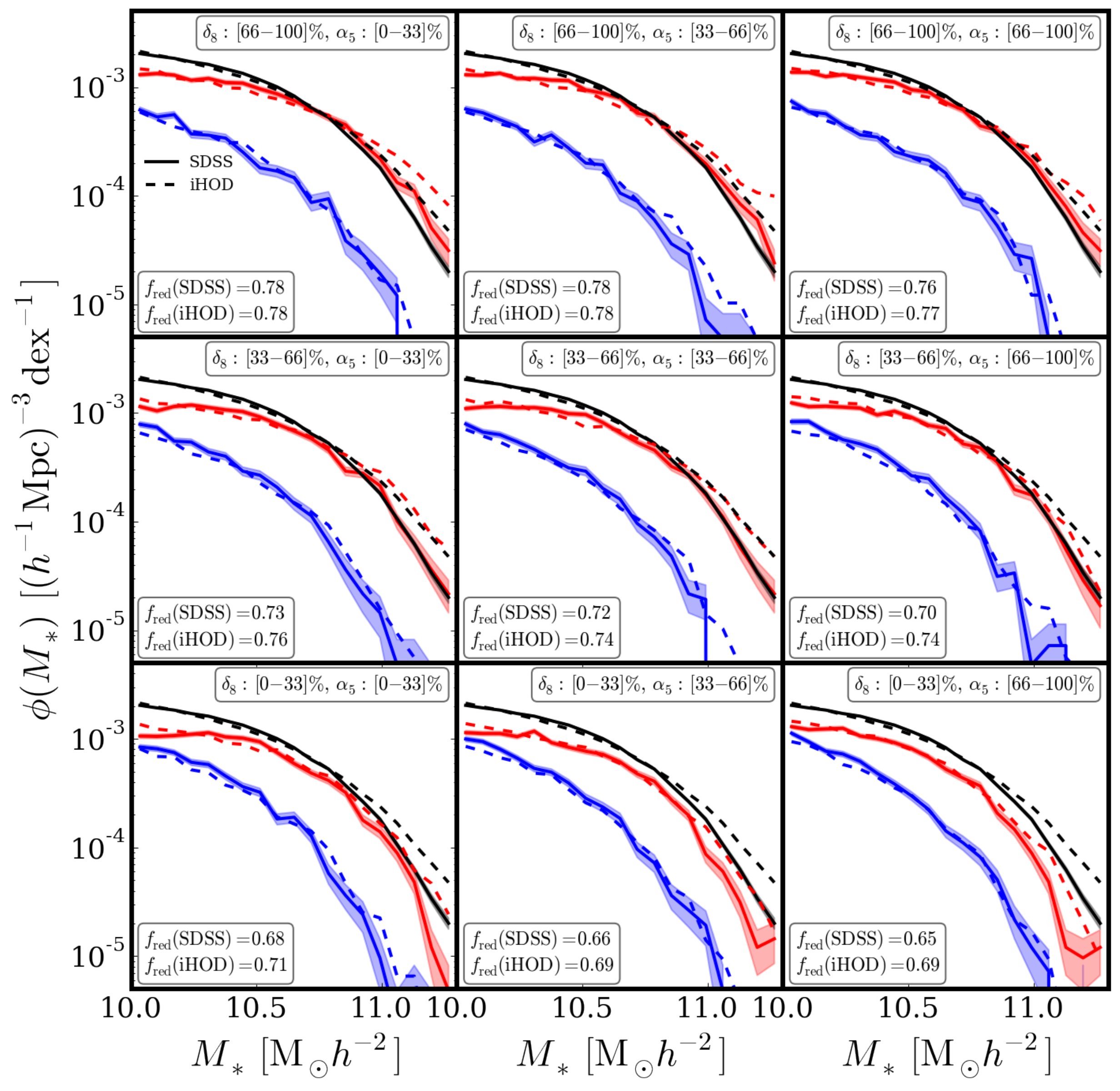}
\caption{Stellar mass functions of red and blue galaxies in 2D bins of $\delta_8$ and $\alpha_5$~(from top to bottom rows: top,
    middle, and bottom $1/3$ in $\delta_8$; from left to right columns: top, middle, and bottom $1/3$ in $\alpha_5$; see the top
    right legend), for
    SDSS~(solid curves with shaded uncertainty bands) and \ihod{}~(dashed curves) samples, respectively. In each panel for each
    sample, the black, red, and blue curves indicate the overall, red, and blue galaxies, respectively, with the overall curve
    normalized to have $1/9$ of the total number density in each sample. The red galaxy fractions in each 2D bin are
    listed in the bottom left corner of each panel. The remarkable agreement of SDSS and \ihod{} in this figure suggests that observed relative abundances of red and blue galaxies do not require any additional cosmic web dependence on top of what is produced through halo mass.}
\label{fig:SMF}
\end{figure*}

Naively, the dependence on $\alpha_5$ seen in the left-hand panel of
Figures~\ref{fig:bias-2d} and \ref{fig:bias-2d-cross} might lead us to conclude that galaxy clustering is indeed
directly influenced by tidal forces within the cosmic web.
In particular, to model the clustering of any stellar
mass~(or luminosity) limited galaxy sample, one usually assume a uniform HOD for that sample across all types
of cosmic web environment. For example, the mean central galaxy occupation can be described as
\begin{equation}
    \left<N_{\mathrm{cen}}\right>(M_h) =
    \frac{1}{2}\left[1-\mathrm{erf}\left(\frac{\log M_h -
        \log M_{\mathrm{min}}}{\sigma_{\log M_h}}\right)\right],
\end{equation}
where $M_{\mathrm{min}}$ corresponds to the halo mass scale that yields $\left<N_{\mathrm{cen}}\right>{=\,}0.5$,
while $\sigma_{\log M_h}$ controls the sharpness of the transition from $\left<N_{\mathrm{cen}}\right>{=\,}0$ to
$1$~\citep{zheng2005}. However, it is plausible that
$M_{\mathrm{min}}$ and $\sigma_{\log M_h}$ are both functions of $\delta_8$ and/or $\alpha_5$~\citep[see][for
an interesting example of a $\delta_8$-dependent HOD]{mcewen2016, wibking2017}. For instance, to boost the clustering of
galaxies with high $\delta_8$ and $\alpha_5$, one could make $M_{\mathrm{min}}$ larger and/or $\sigma_{\log
M_h}$ smaller in the dense filamentary environments than in the field.

If the HOD indeed depends strongly on the cosmic web, one would expect that the \ihod{}\ mock galaxies are
unable to reproduce the observed 2D bias map from SDSS, because the mock catalogue does not include any
dependence on $\delta_8$ or $\alpha_5$. However, the mock 2D bias map shown in the right panel provides a good
match to the observations on the left, qualitatively reproducing all the features discussed above. The mock
bias varies slightly more continuously across adjacent cells than the observed bias due to the better
statistics~(i.e., larger volume) in the mock. There are some discrepancies between the observed and the mock
bias maps within the $\alpha_5$ range between $0$ and $20$ percentiles, echoing the lack of an enhanced bias
at the lowest $\alpha_5$ end in the mock~(see the right panel of Figure~\ref{fig:bias-1d}). However, it is
unclear whether such discrepancies reflect the shortcomings of the \ihod{}\ model in describing the clustering
of galaxies in the extremely low $\alpha_5$ regime, or merely statistical fluctuations that will vanish in
future larger surveys. We will investigate the sources of those discrepancies in more detail in a future paper~(Salcedo et al, in prep.).

Overall, there is good agreement between the data and the \ihod{}\ mock in describing the cosmic web
dependence of galaxy clustering on all scales, as shown by the combination of Figures~\ref{fig:wp-10perc}- \ref{fig:bias-2d-cross}. The only place where \ihod{}\ appears to fail slightly is in predicting the $\alpha$-bias relation to be a little weaker compared to SDSS data as shown in Figure~\ref{fig:bias-1d}. This is encouraging, suggesting that the additional dependence of HOD on the cosmic web is small, as the observed trend of galaxy bias with cosmic web properties can be largely described by the inherent dependence of halo mass function on those properties encoded in the \ihod{} model.

\begin{figure*}
\includegraphics[width=1.0\textwidth]{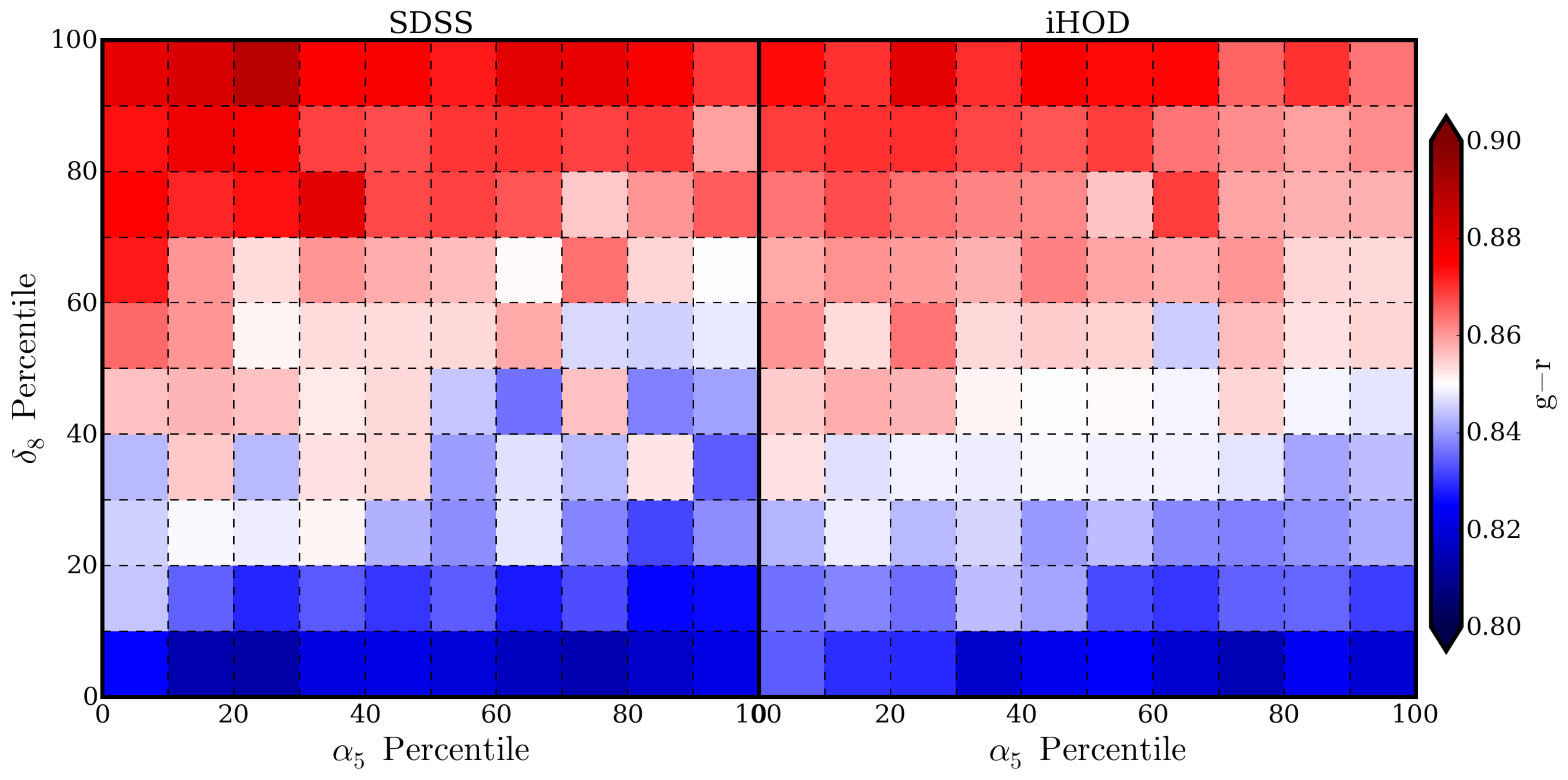}
\caption{Similar to Figure~\ref{fig:bias-2d}, but for the mean $g-r$ colour of galaxies.  The figure
  shows that the mean $g-r$ colour depends both on $\delta_8$ and $\alpha_5$  in both SDSS and
  \ihod{}. The subtle dependence of mean $g-r$ colour on $\alpha_5$ is unlikely to be the evidence
  of tidal dependence of galaxy quenching because it is reproduced by the \ihod{} model, which does
  not include any such effect.}
\label{fig:gcolor}
\end{figure*}

\begin{figure*}
\includegraphics[width=0.8\textwidth]{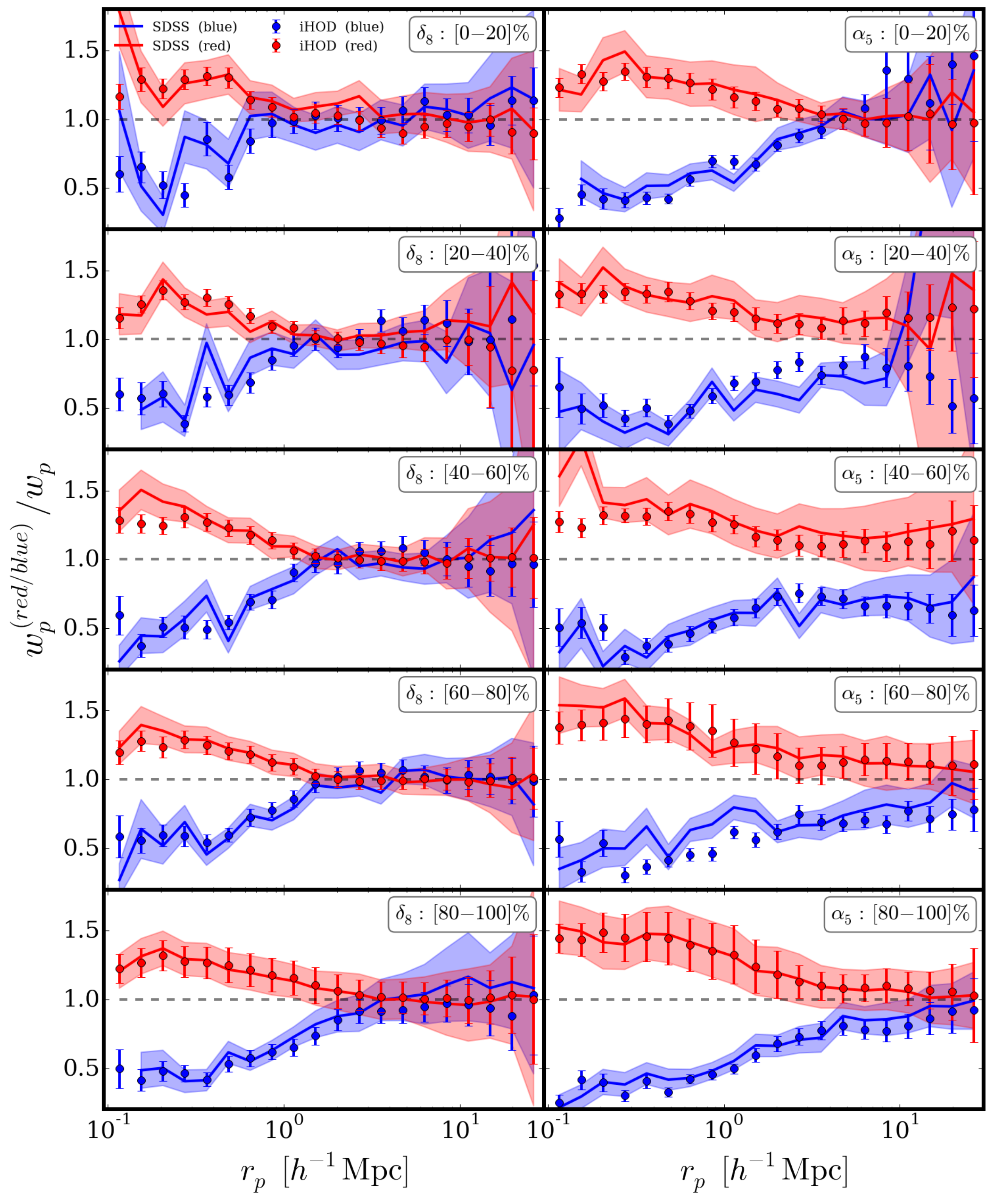}
\caption{The relative clustering of the galaxies in each coloured subsample to that of the overall sample in 20 percentile bins of
    $\delta_8$~(left) and $\alpha_5$~(right), respectively.  In each panel, solid curves with shaded 1-$\sigma$ uncertainty bands
    indicate the SDSS galaxies, and the circles with error bars show results for the \ihod{} galaxies. The relative clustering of each red~(blue)
    subsample is defined as the projected cross-correlation of red~(blue) subsample and the overall sample within that 2D
bin, divided by the projected auto-correlation of the overall sample.  
The two key results shown in this figure are that the characteristic scale at which red and blue
galaxies have the same clustering depends on environment and the large scale relative bias of red
and blue galaxies are the same in all $\delta_8$ bins but differ from each other in some $\alpha_5$ bins.}
\label{fig:wp-color}
\end{figure*}

\subsection{Dependence of galaxy quenching on $\delta_8$ and $\alpha_5$}

We have shown in the above section that, before splitting by any quenching properties, galaxy occupation
statistics seem to be a weak function of the cosmic web environment. However, it is plausible that  the quenching of
galaxies is influenced by the cosmic web despite
the fact that the overall galaxy formation efficiency is largely independent of $\delta_8$ and $\alpha_5$.
For example, the halo gas accretion could be heavily regulated by the tidal anisotropy, so that it is easier
for streams of cold gas to penetrate halo boundaries if they were funnelled through the filaments than in an
isotropic tensor environment, where the `hot mode' accretion usually takes over~\citep{keres2005}.
Therefore, if the quenching of star formation is primarily caused by the shutdown of cold gas accretion,
rather than the removal of the gas reservoir and/or the heating of gas~(through various feedbacks and virial
shocks), we would expect a strong cosmic web dependence of galaxy quenching.

To investigate whether a strong cosmic web quenching exists, we divide the observed galaxies into quenched vs.
star-forming according to their $g{-}r$ colours, and measure the relative abundance and clustering bias of
red~(quenched) vs.  blue~(star-forming) galaxies in different cosmic web environments defined by $\delta_8$
and $\alpha_5$.  Following a similar route as in~\S~\ref{subsec:clustering}, we perform the same analysis over
the \ihod{}\ mock galaxies and compare with the SDSS measurements, in hopes of detecting any cosmic web
dependence of galaxy quenching that is absent from the fiducial halo quenching model of~\citet{zu2017}.

To separate galaxies into red and blue, we use the stellar mass dependent colour split as suggested in
equation 2 of~\cite{zu2016}. We first compare the relative abundances of red vs. blue galaxies between SDSS
and \ihod{} across various cosmic web environments, as shown in Figure~\ref{fig:SMF}. Within each catalogue,
we divide galaxies into nine subsamples based on their $\delta_8$ and $\alpha_5$ percentiles, indicated by the
legend on the top right of each panel. In each panel, we show the red~(red curves) and blue~(blue curves)
galaxy stellar mass functions for the SDSS~(solid) and \ihod{}~(dashed) galaxies, respectively. The red galaxy
fractions of the SDSS and \ihod{}\ subsamples are also listed in the bottom left of each panel.  We show the
jackknife 1-$\sigma$ uncertainties of the SDSS abundances as shaded bands, while the uncertainties for the
\ihod{}\ measurements are much smaller due to better statistics~(not shown here to avoid clutter).

Uniform across all panels, the black solid and dashed curves indicate the overall stellar mass
functions~(SMFs; reduced by a factor of nine for easy comparison) of the SDSS and \ihod{}\ galaxies,
respectively.  The overall SMF of the \ihod{} galaxies agrees very well with that of the SDSS galaxies at
$M_*{<}10^{11} h^{-2}M_{\odot}$.  As emphasized in \citet{zu2015}, this agreement of overall SMFs is not by
construction, as the SMF is a prediction from the best-fitting \ihod{} model, rather than an input to the
constraint. The \ihod{} SMF has a slightly higher amplitude than the observed one at the extreme high mass
end, probably induced by a combination of observational and model systematics. Fortunately, our statistical
results are dominated by galaxies with $M_*{<}10^{11} h^{-2} M_{\odot}$, and are thus robust to any SMF
discrepancy at the very high mass regime.

The relative abundance of red vs. blue galaxies exhibits strong dependence on $\delta_8$ in the SDSS, with the red
galaxy fraction $f_{\mathrm{red}}$ increasing from
${\,\sim\,}68\%$ at low $\delta_8$~(bottom row) to ${\,\sim\,}78\%$ at high
$\delta_8$~(top row), regardless of $\alpha_5$. This $f_{\mathrm{red}}$-$\delta_8$ relation is expected from
the well-known colour-density relation~ \citep{Oelmer1974ApJ...194....1O,Davis1976ApJ...208...13D, Dressler1980ApJ...236..351D}. At fixed $\delta_8$, however, the
observed relative abundance of red vs. blue galaxies shows little variation with $\alpha_5$. The cosmic web
dependence of the colour-split SMFs predicted by the \ihod{} mock shows excellent agreement with that observed
in the SDSS, with the dashed curves closely tracking the solid ones in each panel. Such high level of agreement is
remarkable, because the colour-split SMFs in the \ihod{} model are predicted from the fiducial halo quenching
model, which is constrained solely by the spatial clustering and weak lensing of the red vs. blue galaxies in
SDSS. Since the cosmic web dependences of the colour-split SMFs in the \ihod{} mock are entirely inherited from
the cosmic web dependence of the halo mass functions via halo quenching, this agreement suggests that no
cosmic web quenching is required to reproduce the observed relative abundance of red vs.  blue galaxies across
different cosmic web environments.

An alternative quantity for statistically characterizing the quenched level of a galaxy sample is the average
colour of that sample $\left<g{-}r\right>$, which has the advantage of not relying on a somewhat arbitrary
colour split to divide galaxies into red and blue. The disadvantage is that the mock galaxy colours, if
predicted by semi-analytic models or hydrodynamic simulations, usually have difficulties reproducing the
observed colour distributions to very high precision~\citep{2015MNRAS.451.2663H, Sales2015MNRAS.447L...6S,
    Bray2016MNRAS.455..185B}. However, thanks to the colour assignment scheme implemented
in~\citet{zu2017}, at any fixed $M_*$ the mock colours follow a double-Gaussian distribution that is inferred
from a volume-limited sample in the SDSS. Therefore, we can directly compare the cosmic web dependence
of $\left<g{-}r\right>$ predicted by the \ihod{} halo quenching model to that observed in the SDSS.

We compute $\left<g{-}r\right>$ as a 2D function of $\delta_8$ and $\alpha_5$, using the same subsamples
defined earlier for the 2D relative bias comparison.  Figure~\ref{fig:gcolor} shows the result for the SDSS
galaxies on the left and \ihod{} mock on the right.  The mean $g{-}r$ colour of the full \ihod{} catalogue is the
same as that in the SDSS~($0.85$; shown as white colour on the 2D colour map).  Consistent with our findings in
Figure~\ref{fig:SMF}, the average colour depends primarily on $\delta_8$ in both catalogues, with the
reddest~(bluest) population living in regions with the top~(bottom) $10\%$ $\delta_8$. However, the left panel
of Figure~\ref{fig:gcolor} reveals a subtle $\alpha_5$ dependence that was indiscernible in
Figure~\ref{fig:SMF} --- at fixed narrow range of $\delta_8$, galaxies that live in high-$\alpha_5$ regions
appear to be slightly bluer than those live in low-$\alpha_5$ regions. Note that this $\alpha_5$ dependence
reflects an overall reduction in the quenched fraction, rather than any relative decrease of the
high-$M_*$~(thus redder) galaxies in high-$\alpha_5$ environments~(which is not seen in Figure~\ref{fig:SMF}).

One might think that the physical cause of this subtle $\alpha_5$ dependence is related to the impact of tidal
anisotropies on galaxy quenching. However, in the right panel of Figure~\ref{fig:gcolor}, the cosmic web
dependence of the mean colour predicted by the fiducial halo quenching model shows very good consistency with
the observations on the left, including both the strong $\delta_8$ dependence and the subtle variation with
$\alpha_5$. As a result, this agreement shown in Figure~\ref{fig:gcolor} between SDSS and \ihod{} suggests
that this intriguing $\alpha_5$-dependence of $\left<g{-}r\right>$ is unlikely to be evidence of direct tidal
impact on galaxy quenching, but is again an imprint left by halo quenching via the cosmic web modulation of halo
mass function. 

A similar dependence of mean colour on tidal environment was found by
\citet{Yan2013MNRAS.430.3432Y} (see their Figures 3 \& 4). They showed that such a dependence was
mostly due to the choice of smoothing scale, and that it will disappear if one uses an adaptive smoothing
that maximizes the density-colour relation. We have also looked at the effect of smoothing scale on
our mean colour with over-density and tidal anisotropy. We found that using a different smoothing
scale does not affect the mean colour for regions with over-density below the 50th percentile. We
observed that as we reduce our smoothing scale, the colour of high over-density and low tidal
anisotropy (top left part) reduces at the same time as the mean colour of high over-density and
high tidal anisotropy (top right) increases. This might happen because we are working in apparent
redshift space and hence while using smaller smoothing we become more sensitive to the distortion in
density introduced by peculiar velocity.  Hence, clusters (which should have an isotropic tidal
environment) appear more and more anisotropic (filamentary), leading to a high value of
$\alpha$. Therefore if one is working in apparent redshift space it will be important for the
mocks to reproduce the peculiar velocity of galaxies, 
in order to match such colour distributions for smaller smoothing scales.

Finally, we examine the relative clustering of red vs. blue galaxies as a function of $\delta_8$~(left) and of
$\alpha_5$~(right) in Figure~\ref{fig:wp-color}.  In each column, we divide galaxies in each of the two
catalogues into five equal-size bins according to their $\delta_8$ or $\alpha_5$. In each $20$ percentile bin,
we compute the relative clustering of a coloured sample as the ratio of $w_p^{\mathrm{color}}$ and $w_p$, where
$w_p^{\mathrm{color}}$ is the cross-correlation between the red/blue galaxies and all galaxies in that bin,
and $w_p$ is the auto-correlation of all galaxies in that bin. In each panel, solid curves with shaded bands
are the SDSS measurements and $1$-$\sigma$ uncertainties, while the solid circles with error bars are from the
\ihod{} mock.

At fixed $\delta_8$~(left panels), the red and blue SDSS galaxies show very similar clustering biases on large
scales. This can be understood by examining Figure~\ref{fig:bias-2d},~\ref{fig:SMF}, and~\ref{fig:gcolor}
simultaneously --- although the large-scale bias varies with $\alpha_5$ at fixed $\delta_8$, the red galaxy
fraction and average galaxy colour vary much less so with $\alpha_5$, indicating that the red and blue galaxies
are not segregated into regions of different biases. However, the clustering amplitudes of the red and blue
galaxies exhibit strong discrepancies on small scales, and the characteristic scale at which the two start
deviating increases monotonically with $\delta_8$, from ${\,\sim\,}0.6\mpcoh$~(top left) to
$3\mpcoh$~(bottom left). In the lowest $\delta_8$ bin, this characteristic quenching scale roughly
corresponds to the size of the most massive haloes in that under-dense environment, whereas in the highest
$\delta_8$ bin it grows much larger than the size of a single massive halo. The large characteristic quenching
scale in dense regions is consistent with the so-called `2-halo conformity' phenomenon~\citep{kauffmann13},
i.e., the coherent quenching of galaxies within the same large-scale overdensity environment.  However, the
relative clustering of the red vs. blue galaxies predicted by the \ihod{} halo quenching model~(solid circles)
is in excellent agreement with the measurements from SDSS in all $\delta_8$ environments, including the
small-scale discrepancy between the clustering amplitudes of the two colours, as well as the characteristic
quenching scale as a function of $\delta_8$. This agreement echoes the findings in \citet{zu2017}, that the
observed 2-halo conformity effect can be largely explained by the simple halo quenching, mediated by the
environmental dependence of halo mass functions~\citep[see also][]{tinker17, sin17, calderon2017}.

Although we have seen that red and blue galaxies have identical large-scale bias when considering subsamples
at constant $\delta_8$ this is not the case when we split by $\alpha_5$. In the right-hand panels
of Figure~\ref{fig:wp-color}, we see that the red SDSS galaxies exhibit stronger large-scale
clustering bias than the blue ones at high $\alpha_5$~(bottom three panels on the right), but have a
comparable bias at low $\alpha_5$~(top two panels on the right).  This finding can be understood as follows.
Figure~\ref{fig:bias-2d} shows that the large-scale bias increases monotonically with $\delta_8$ at high
$\alpha_5$, while according to Figure~\ref{fig:gcolor} the galaxies are progressively more quenched at higher
$\delta_8$ at any fixed $\alpha_5$.  Therefore, we expect the red galaxies to have a stronger large-scale bias
than the blue ones at high $\alpha_5$.  However, the large-scale bias changes non-monotonically with
increasing $\delta_8$ at low $\alpha_5$, so that the low and high-$\delta_8$ galaxies have comparable
large-scale biases~(see Figure~\ref{fig:bias-2d}).  By the same token, we expect the bias-colour trend to be
reversed at low $\alpha_5$ and that the blue galaxies have a similar large-scale bias as the red ones.  On
scales below $5\mpcoh$, the red galaxies clustered more strongly than the blue ones across all
$\alpha_5$ environments. Similar to the left panels, the observed $\alpha_5$ and scale dependences of the
relative clustering of two colours are closely reproduced by the predictions from the \ihod{} quenching model.

The combination of Figures~\ref{fig:SMF}, ~\ref{fig:gcolor} \&~\ref{fig:wp-color} demonstrates that, although
the observed galaxy quenching properties~(including the relative abundance and clustering of the red vs. blue
galaxies) have a complicated apparent dependence on cosmic web, those dependences can be entirely explained by
the fiducial halo quenching model in \ihod{}, constrained solely by the spatial clustering and the weak
lensing of red and blue galaxies in the SDSS. The surprising success of such a simple quenching model, described
by merely four parameters, strongly suggests that the galaxy quenching process is primarily driven by halo
mass, and the large-scale~(${>}5\mpcoh$) tidal environment has little direct influence on the colour
transformation of galaxies above $10^{10}\,h^{-2}M_{\odot}$.

\section{Summary and Discussion}
\label{sec:summary}

We have measured the dependence of galaxy clustering and quenching on the cosmic web properties in the SDSS,
characterized by the combination of galaxy overdensity~($\delta_8$) and tidal anisotropy~($\alpha_5$) at each
galaxy position. We have developed an improved method to measure the tidal tensor field from the observed
galaxy distribution in redshift surveys, by applying more accurate density estimation scheme using tessellations. We always work with the apparent redshift-space tidal field estimated using exactly same method for SDSS galaxies and mock galaxies.
In order to reveal any {\it direct} cosmic web effect on galaxy formation, we compare the
observed dependences in the SDSS to those measured from an HOD galaxy mock catalogue built from the \ihod{}
fiducial halo quenching model of~\citet{zu2015, zu2016, zu2017}. The galaxy-cosmic web relation in the \ihod{}
mock is solely derived from the cosmic web modulation of the underlying halo mass function, as the stellar
mass, occupation statistics, and optical colour of the mock galaxies are determined only by the mass of the
host haloes.

The projected correlation function~($w_p$) of galaxies shows complicated $\delta_8$ and $\alpha_5$ dependences
on both the small and large scales. In particular, the large-scale clustering bias of galaxies exhibits a
characteristic dependence on $\delta_8$ as well as $\alpha_5$, with the 1D bias functions having 
their lowest absolute values around the $30$ percentile of $\delta_8$ and $\alpha_5$, respectively.
We investigated the origin of such
characteristic dependences by examining the 2D galaxy bias function on the $\delta_8$-$\alpha_5$ diagram, which
reveals a diagonal band of low bias regions at intermediate values of $\delta_8$ and $\alpha_5$. 

However,
despite the high level of complexities in the observed cosmic web dependences of $w_p$, they can all be very closely
reproduced by our simple \ihod{} galaxy mock with no inbuilt cosmic web effects on galaxy clustering.
We constrain the rms of slope of any additional dependence of relative bias on $\alpha_5$ to be
$\sigma_{m_{\alpha_5}}=0.30$. Therefore any model producing $m_{\alpha_5}>0.9$ is ruled out for our
sample at the $3\sigma$ significance level.
We obtain  $m_{\alpha_5}=0.69 \pm 0.30$ for our \ihod{} mock galaxy catalogue.
This could be taken as providing some weak evidence of cosmic web effects on the spatial clustering of galaxies, 
implying that the stellar-to-halo mass relation and the HOD of satellite galaxies might depend on the
local tidal field. But this could as well be due the fact that we work in  apparent redshift space and the peculiar velocity in the mocks are not modelled accurately. Some preliminary test shows that strong difference in peculiar velocity could produce such a deviation.

As a statistical measure of quenching, the mean $g{-}r$ colour of SDSS galaxies is a strong function of
$\delta_8$, with some subtle but non-trivial dependence on $\alpha_5$ at fixed $\delta_8$. We also examined
the relative clustering of red vs. blue galaxies across different $\delta_8$ and $\alpha_5$ regions, which
probes the spatial segregation of the two colours in each region. Remarkably, the observed dependences of
galaxy mean colour and colour segregation are in excellent agreement with the predictions from the \ihod{}
fiducial halo quenching model, including the subtle mean colour dependence on $\alpha_5$ and the
characteristic scale at which the relative clustering starts deviating from unity. This agreement is
encouraging news for the theoretical interpretation of galaxy quenching in the halo model, as the tidal tensor
field can be largely ignored when modelling the colour transformation of galaxies.

Our analysis could be affected by a few systematic uncertainties in the measurement. In particular, our
compact characterization of the cosmic web could be overly simplistic, and therefore may not capture the full
properties of the tidal tensor field. For instance, the tidal field may operate on galaxy formation over a
smaller distance scale than $5\mpcoh$, the one we used for defining $\alpha_5$. However, we expect a strong
correlation between the tidal anisotropies defined at various different scales below the correlation length of
galaxy clusters~\citep{croft1997}. Therefore, the use of
$\alpha_5$ may not be optimal, but the effect of tidal anisotropy should nonetheless show up in our analysis
if a strong cosmic web effect on clustering and/or quenching indeed exists. Another potential source of
systematic error is the modelling of peculiar velocities in the mock. We have implemented the best-fitting
velocity bias model of \citet{guo2015b}, which assumes that the relative velocities of central and satellite
galaxies follow Gaussian distributions within each halo. However, the velocity distribution of galaxies has
some level of anisotropy that is likely to be correlated with the tidal anisotropy outside the halo, especially in
massive clusters~\citep{zu2013}. We expect that the problem can be at least partially resolved in the future
by directly using the velocities of subhaloes, which would require a simulation of higher resolution than
\texttt{Bolshoi}.

Theoretically speaking, the success of the simple \ihod\ halo quenching model further strengthens
the argument that the halo mass plays the primary role in shaping the galaxy content inside individual dark
matter haloes.
Similarly, the lack of strong cosmic web effects on the colour
transformation of galaxies implies that the galaxy quenching is a relatively local process that is contained
within the boundary of haloes~\citep{baxter2017}, and that the tidal modulation of the large-scale cold gas
accretion is unlikely to be the controlling factor triggering galaxy quenching. An exciting prospect is to apply a
similar analysis to the cosmic web at higher redshifts~\citep[e.g., DESI;][]{desi2016}, as hydrodynamic simulations predict that the
temperature distribution of the gas accretion becomes more bimodal with increasing redshift~\citep{keres2005}.

While this work was being completed, we became aware of a related analysis by \cite{paranjape2018}. These authors have analysed the tidal environment of SDSS galaxies using a different set of selection criteria, analysis techniques, and definition of tidal anisotropy, and focused exclusively on the dependence of galaxy clustering on tidal anisotropy. Encouragingly, the results of our two independent analyses on the cosmic web dependences of galaxy clustering, wherever they can be compared, are qualitatively consistent.

\section*{Acknowledgments}

We thank David H. Weinberg for helpful discussions. We would like to
thank Uros Seljak, Sukhdeep Singh and Hung-Jin Huang for comments and
discussion on the bias measurement and the importance of
cross-correlations. We would also like to thank Simon White and
Katarina Kraljic for useful comments on an earlier version of the
draft. SA and JAP are supported by the European Research Council
through the COSFORM Research Grant (\#670193).  YZ is supported by a
CCAPP fellowship and the Thousand Talents Program of China.  We thank
the Bolshoi team for making their simulations publicly available. This
research has made use of the NASA/IPAC Extragalactic Database (NED)
which is operated by the Jet Propulsion Laboratory, California
Institute of Technology, under contract with the National Aeronautics
and Space Administration. This research has made use of NASA's
Astrophysics Data System.

%%%%% Bibliography %%%%%%%%%%%%%%%%%%%%%%%%%%%%%%%%%%%%%%%%%%%%%%%%%%%%%%%%%%%%

\bibliography{Master_Shadab}
\bibliographystyle{mnras}

\label{lastpage}

\end{document}